\documentclass[%
 reprint,
 amsmath,amssymb,
 aps,
 noeprint,
 prx,
]{revtex4-2}

\usepackage{graphicx}% Include figure files
\usepackage{xcolor}
\usepackage{dcolumn}% Align table columns on the decimal point
\usepackage{amsmath}
\usepackage[ruled,vlined]{algorithm2e}
\usepackage{braket}
\usepackage{mathtools}
\usepackage{bm}% bold math
\usepackage[colorlinks,linkcolor=blue,citecolor=red,urlcolor=blue]{hyperref}% add hyper capabilities
\usepackage[normalem]{ulem}

\newif\ifptitle
\newif\ifpnumber
\newcounter{para}

\ptitletrue  % comment this line to hide paragraph titles
\pnumbertrue  % comment this line to hide paragraph numbers

\newcommand{\HarvardPhysics}{Department of Physics, Harvard University, Cambridge, MA 02138, USA}

\usepackage{lipsum} % for filler text

\begin{document}

\title{
Enhancing Quantum Memory Lifetime with Measurement-Free Local Error Correction and Reinforcement Learning
}

\author{Mincheol Park}
\email{mincheol\_park@g.harvard.edu}
\affiliation{\HarvardPhysics}

\author{Nishad Maskara}
\affiliation{\HarvardPhysics}

\author{Marcin Kalinowski}
\affiliation{\HarvardPhysics}

\author{Mikhail D. Lukin}
\email{lukin@physics.harvard.edu}
\affiliation{\HarvardPhysics}

\date{\today}  % It is always \today, today,
               % but any date may be specified

\begin{abstract}
    Reliable quantum computation requires systematic identification and correction of errors that occur and accumulate in quantum hardware.
    To diagnose and correct such errors, standard quantum error-correcting protocols utilize \emph{global} error information across the system obtained by mid-circuit readout of ancillary qubits.
    We investigate circuit-level error-correcting protocols that are measurement-free and based on \emph{local} error information.
    Such a local error correction (LEC) circuit consists of faulty multi-qubit gates to perform both syndrome extraction and ancilla-controlled error removal.
    We develop and implement a reinforcement learning framework that takes a fixed set of faulty gates as inputs and outputs an optimized LEC circuit.
    To evaluate this approach, we quantitatively characterize an extension of logical qubit lifetime by a noisy LEC circuit.
    For the 2D classical Ising model and 4D toric code, our optimized LEC circuit performs better at extending a memory lifetime compared to a conventional LEC circuit based on Toom's rule in a sub-threshold gate error regime.
    We further show that such circuits can be used to reduce the rate of mid-circuit readouts to preserve a 2D toric code memory.
    Finally, we discuss the application of the LEC protocol on dissipative preparation of quantum states with topological phases.
\end{abstract}

\maketitle

% \tableofcontents

\section{Introduction}

% \ptitle{Standard/Global QEC}

Quantum error correction (QEC) protects encoded quantum information from experimental noise and is critical to realizing the full computation power of quantum processors~\cite{nielsen00}.
By embedding logical information into QEC codes, errors can be systematically identified and corrected~\cite{PhysRevA.52.R2493}.
In recent years, experimental progress has been rapid with realization of small codes and simple correction procedures being explored across various quantum hardware platforms, including the 2D triangular color code \cite{PhysRevX.11.041058} on trapped ions, the 2D surface code \cite{acharya2022suppressing} on superconducting qubits, and the 2D toric code \cite{Bluvstein_2022} and 3D color codes \cite{Bluvstein_2023} on neutral-atom arrays.
However, the ultimate path to efficient and scalable error correction in large systems is far from clear.
As such, a variety of approaches for reducing hardware requirements should be considered.

% \bigskip\ptitle{Mid-circuit readout cost}

Stabilizer codes are a particularly important class, as they come equipped with efficient circuits for identifying and correcting errors without affecting the encoded information~\cite{PhysRevA.54.1862, gottesman1997stabilizer}.
One approach to implement error correction with these codes is to use ancillary qubits and readouts to extract stabilizer eigenvalues.
Then, stabilizer information is aggregated \emph{globally} using a classical decoding algorithm, which identifies likely errors and implements corresponding recovery operations on the data qubits comprising the codes.
These include decoders based on the minimum-weight perfect matching algorithm~\cite{Dennis_2002, wang2009threshold, PhysRevA.86.032324, PhysRevA.89.012317, Nickerson2019analysingcorrelated, higgott2021pymatching, PhysRevA.108.022401, higgott2023sparseblossomcorrectingmillion}, machine learning~\cite{PhysRevA.99.052351, chinni2019neural, Varsamopoulos_2020, bhoumik2021efficient, egorov2023end, bausch2023learning, Gicev_2023, fu2023benchmarking, hall2023artificial, wang2023transformerqec}, or other approaches~\cite{PhysRevA.90.032326, Criger2018multipathsummation, Delfosse2021almostlineartime, PhysRevResearch.2.033042}.
However, in principle, the correction procedure could be performed without intermediate communication with a classical processor, using coherent multi-qubit controlled gates and dissipation~\cite{PhysRevLett.127.050501, Deist_2022, PhysRevX.13.041034, PhysRevX.13.041051, Singh_2023, Evered_2023, PhysRevX.13.041035}.

% \bigskip\ptitle{Past passive error correction/LEC proposals}

More specifically, there are measurement-free QEC protocols utilizing only \emph{local} operations in its decoding procedure.
These local error correction (LEC) strategies have received considerable theoretical attention, in particular in higher dimensions.
Indeed, in dimensions $D \geq 4$, local cellular automata-based decoders have reported thresholds under phenomenological noise models, indicating the possibility of self-correcting quantum memories~\cite{10.5555/3179532.3179533, Herold_2015, Kubica_2019, Vasmer_2021, Scruby_2023, ray2023protecting}.
Moreover, efforts have extended into optimizing local decoding protocols by training convolutional neural networks \cite{Breuckmann_2018, Meinerz_2022}, 
and practical implementation and application of local decoders in near-term noisy quantum systems \cite{Smith_2023, heussen2023measurementfree, guedes2023quantum, PhysRevA.108.062426, veroni2024optimized, zhu2024passiveerrorcorrectionqubitoscillator}.
Nevertheless, designing LEC circuits with a restricted faulty multi-qubit gate set is a challenging task.
For this purpose, it is desirable to develop a systematic optimization framework that enables such circuit design.

% \bigskip\ptitle{Past RL in quantum science/engineering}

Reinforcement learning (RL) is a powerful framework for optimizing sequences of operations to perform a particular task~\cite{10.5555/1622737.1622748}.
As such, RL has been applied recently in quantum science and engineering~\cite{PhysRevA.107.010101} to develop both quantum error-correcting codes \cite{Nautrup_2019, su2023discovery, olle2024simultaneous} as well as decoders~\cite{Andreasson_2019, Sweke_2020, Fitzek_2020, matekole2022decoding}.
RL was also utilized to optimize noisy quantum circuits or controls that perform variational quantum algorithms~\cite{patel2024curriculum, kundu2024reinforcement, chen2024qubitwise}, logical quantum gates~\cite{ivanovarohling2024discovery}, and state preparation~\cite{Menegasso_Pires_2021, Baba_2023, Li_2023, kolle2024reinforcement}.
For the state-of-the-art or near-term quantum hardware with few data qubits, RL was applied to optimize quantum circuits that preserve quantum memory \cite{F_sel_2018} and prepare logical states \cite{zen2024quantum}.

In this work, we optimize circuit-level LEC protocols that consist of faulty multi-qubit gates --- to perform both syndrome extraction and ancilla-controlled error removal --- by developing an RL framework.
In particular, our RL framework takes a fixed set of faulty gates as inputs and outputs an LEC circuit, with optimized length and layout.
By performing extensive circuit-level simulations, we show these circuits successfully reduce the error density when gate fidelities are high enough.
We then characterize the resulting extension in the lifetime of a memory encoded in three finite-sized systems: 2D toric code, 2D Ising model, and 4D toric code.
Finally, we explore two near-term applications of this scheme: (1) reducing the rate of mid-circuit readouts required to preserve quantum memory and (2) dissipatively preparing quantum states of topological phases.

This paper is organized as follows: we begin in Section \ref{sec: overview} by outlining the key insights and main results of this work.
Section \ref{sec: II} provides a detailed analysis of a circuit-level LEC scheme.
In particular, we present an implementation of the basic components of the LEC circuits followed by characterization and resolution of the error patterns that limit conventional LEC circuits.
Then, Section \ref{sec: III} discusses a setting for the RL framework and its optimization of circuit depth and layout within each code.
Section \ref{sec: IV} presents a quantitative characterization of memory lifetime extension by LEC circuits and a comparison between RL-optimized and conventional LEC circuits.
Furthermore, Section \ref{sec: V} proposes two near-term applications of this scheme: reducing the rate of mid-circuit readouts required to preserve quantum memory in 2D toric code and dissipatively preparing quantum states of topological phases.
% following the considerations of the experimental implementation of the LEC scheme in Section \ref{sec: VA},
%
Finally, we present conclusions and outlook in Section \ref{sec: Discussion}.

\section{Overview of Main Results} \label{sec: overview}

% \bigskip\ptitle{Overview I: RL-optimized LEC Circuit}

\begin{figure*}[t]
    \centering
    \includegraphics[width = \textwidth]{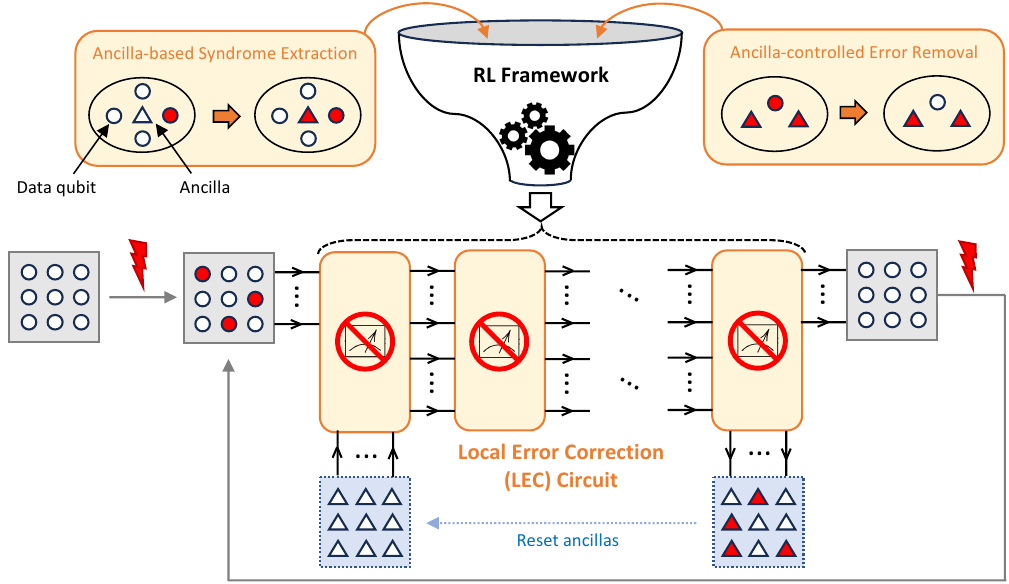}
    \caption{
    Overview of circuit-level local error correction (LEC) protocol and the reinforcement learning (RL) optimization framework.
    The LEC circuit is a sequence of local coherent quantum operations --- not including a mid-circuit projective readout of any qubit --- applied to data qubits (circle) and ancillas (triangle).
    It takes data qubits with errors (colored red) and initialized ancillas as input and moves errors of data qubits into ancillas.
    This local decoding can be repeated over multiple cycles with a finite number of ancillas by resetting them into the initialized ones.
    A reinforcement learning (RL) framework constructs an optimized LEC circuit by combining two types of ``building block'' operations, followed by corresponding coherent and independent reconfiguration of qubits.
    First, \emph{ancilla-based syndrome extraction} operations enable the ancillas to measure and store syndrome information that indicates the existence of errors on neighboring data qubits.
    On the other hand, \emph{ancilla-controlled error removal} operations provide local feedback operation controlled by ancillas onto data qubits to correct their errors.
    All these operations can be achieved experimentally within the state-of-the-art quantum processors, including the Rydberg atom arrays (see Appendix \ref{appendix: exp}).
    }
    \label{fig:1_Overview}
\end{figure*}

We focus on a circuit-level, measurement-free LEC protocol, composed of faulty multi-qubit gates.
As illustrated in Fig. \ref{fig:1_Overview}, the circuits are constructed from two types of basic local quantum operations: (1) ancilla-based syndrome extraction with a series of two-qubit gates and (2) ancilla-controlled error removal with three-qubit gates.
We focus on three finite-sized error-correcting codes --- the 2D toric code, classical 2D Ising model, and 4D toric code --- and start by studying the performance of existing LEC approaches, such as nearest-neighbor matching in 2D toric code~\cite{Smith_2023} or Toom's rule in 2D Ising model and 4D toric code~\cite{10.5555/3179532.3179533}.
We then develop and apply a new RL framework to optimize LEC circuits and improve the overall performance in all three cases.
Finally, we provide a detailed and intuitive understanding of how the optimization achieves improved performance.

%% Fig. 1: Overview

% \bigskip\ptitle{Overview II: Why RL works?}

We find that the existing LEC methods are limited by specific low-weight error patterns, which are not removed by the local error removal operations comprising the circuit, as also noted by previous literature~\cite{Breuckmann_2018, Kubica_2019}.
To address this, we introduce additional longer-range operations that are capable of removing some of these error patterns and thus increase the weight of the minimum uncorrectable error in each code.
The RL framework, as shown in Fig. \ref{fig:2A_SummaryRL}, assembles the operations into optimized LEC circuits that prevent the accumulation of higher-weight error patterns on average over multiple rounds of their application.
In general, since we are considering a circuit-level noise model, adding additional gates also increases the number of errors.
Importantly, we find that the RL can balance this cost, with the benefits arising from removing such errors: as gates get noisier, the optimization prefers shorter and simpler LEC circuits to minimize errors.
In contrast, when we consider gates with higher fidelity, the RL procedure uses more complex decoding procedures to further improve performance.

% \bigskip\ptitle{Overview III: LEC on memory lifetime}

We illustrate the utility of RL-optimized LEC circuits by showing how they can extend the lifetime of quantum memory in a sub-threshold gate error regime.
In each LEC round, we introduce ``ambient'' errors, which model noises arising from other faulty operations including idling or transversal logical gates~\cite{Bluvstein_2023, ryananderson2024highfidelity}, with $p_{\rm{amb}}$ error rate.
Then, we apply a full LEC circuit that consists of faulty multi-qubit controlled gates with $p_{\rm{gate}}$ error rate.
We define an average logical qubit lifetime $T$ to be an average number of LEC rounds before encountering the first failure in decoding.
Simulating different values of $p_{\rm{amb}}$, we observe a scaling of the error suppression consistent with a power law dependence upon $T$ and $p_{\rm{amb}}$ (see Fig. \ref{fig:3C-1_FittingDeff}).
%{\color{red} Fig.xy}.
%
To quantify this effect, we introduce the notion of an \emph{effective code distance} $D_{\rm{eff}}$ as a function of a given LEC circuit and $p_{\rm{gate}}$ such that $T \propto (p_{\rm{amb}})^{-D_{\rm{eff}}}$.
$D_{\rm{eff}}$ is related to the ability of a noisy decoding circuit to remove (on average) error patterns of a certain weight.
In the sub-threshold $p_{\rm{gate}}$ regime, we numerically show that RL-optimized LEC circuits have higher $D_{\rm{eff}}$ compared to conventional LEC circuits in each code and are thus more effective in removing errors from noisy states (see Fig. \ref{fig:3C-2_ComparingDeff}).

% \bigskip\ptitle{Overview IV: LEC applications}

Finally, we explore two potential applications of our LEC method.
First, we explore how RL-optimized LEC circuits can be interleaved with standard (global) decoding, to reduce the number of mid-circuit measurements required to achieve a target performance, which could be used to e.g. reduce cycle times in logical processors.
For example, we show that using LEC circuits with high-fidelity multi-qubit gates can reduce the number of required mid-circuit readouts (see Fig. \ref{fig:4A_ReducingGlobal}).
Second, noting that ideas from LEC have also been used to rigorously verify non-trivial, topological phases of matter in the presence of experimental noises~\cite{liu2023dissipative, Cong_2024}, we show that our LEC scheme can similarly be applied in the dissipative preparation of topological phases with a given Hamiltonian.

\section{Circuit-level LEC Scheme} \label{sec: II}

In this section, we overview our family of circuit-level LEC schemes in 2D toric code, classical 2D Ising model, and 4D toric code by providing an explicit translation into a quantum circuit with multi-qubit gates from the local decoding rule~\cite{toom1974,toom1980stable, 10.5555/3179532.3179533, Smith_2023}.
Section \ref{sec: IIA} discusses how to implement the basic components of the LEC circuit.
Section \ref{sec: IIB} characterizes the error patterns that limit conventional LEC circuits and then introduces the hierarchically extended set of error removal operations to address this limitation of the circuit-level LEC scheme.
To provide relevant details, we include the considerations of the experimental implementation of the LEC scheme in Appendix \ref{appendix: exp} and implementation of a code base outlined in this section in Appendix \ref{appendix: QECC-implementation}.

% \bigskip\ptitle{Intro to definition \& role of LEC circuit}

LEC circuits are constructed as sequences of \emph{local} coherent quantum operations between data qubits and ancillas (see Fig.~\ref{fig:1_Overview}).
Importantly, there is no need for readout and classical processing in the circuit.
Conceptually, the circuit takes noisy data qubits and noise-free ancillas as input and reduces errors on the data qubits by moving them to the ancillas through entangling operations.
Thus, the LEC circuit effectively moves entropy from data qubits onto the ancillas.
In our formulation, we distinguish errors arising from the LEC correction and errors arising from other sources, such as idling errors or other logical-circuit elements.
As such, after each application of the LEC circuit, a new set of i.i.d errors are applied to data qubits.
The ancillas are then reset or discarded and the process is repeated, forming a \emph{cycle} of LEC.

% \bigskip\ptitle{Intro to RL framework \& its components}

As depicted in Fig. \ref{fig:1_Overview}, the RL framework functions as a ``funnel'' that takes the set of available local operations as input and produces a sequence composed of these elements, with optimized length and layout.
We discuss the RL framework in more detail in Section \ref{sec: III}.
The input operations naturally decompose into two functions:
\begin{itemize}
    \item \emph{Ancilla-based syndrome extraction} by encoding stabilizer operators onto ancillas without readout,
    \item \emph{Ancilla-controlled error removal} by applying local feedback controlled by ancillas onto data qubits.
\end{itemize}
Each of these operations requires applying multi-qubit gates to the system \emph{in parallel} following a desired layout.

\subsection{Circuit-level LEC operations in each code}\label{sec: IIA}

% \ptitle{Definition of stabilizers in each code}

\begin{figure}[t]
    \centering
    \includegraphics[width = 0.48 \textwidth]{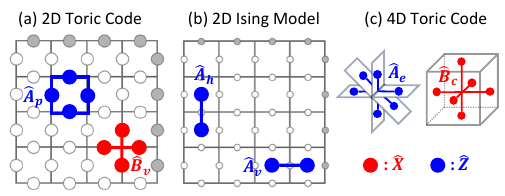}
    \caption{
    Data qubits and stabilizers for each error-correcting code.
    Pauli $\hat{X}$ and $\hat{Z}$ operators on qubits at corresponding lattice positions are colored red and blue, respectively.
    Also, grey qubits denote a periodic boundary condition of a lattice.
    (a) In the 2D toric code lattice, data qubits live at edges.
    Stabilizer $\hat{A}_p$ and $\hat{B}_v$ are defined at each plaquette $p$ and vertex $v$, respectively, by the tensor product of four $\hat{Z}s$ and four $\hat{X}$s, respectively, on adjacent edges.
    (b) In the 2D Ising model lattice, data qubits live at plaquettes.
    Stabilizer $\hat{A}_v$ and $\hat{A}_h$ are defined at each edge by the tensor products of two $\hat{Z}$s on adjacent plaquettes.
    (c) In 4D toric code lattice, data qubits live at faces.
    Stabilizer $\hat{A}_e$ and $\hat{B}_c$ are defined at each edge $e$ and cube $c$, respectively, by the tensor products of six $\hat{Z}$s and $\hat{X}$s, respectively, on adjacent faces.
    }
    \label{fig:1A-1_StabilizerDefinition}
\end{figure}

We will use three models to develop and test our framework, illustrated in Fig. \ref{fig:1A-1_StabilizerDefinition}.
The 2D toric code is a canonical quantum error-correcting code, which has been studied extensively~\cite{Dennis_2002, wang2009threshold}.
This code, having point-like syndrome excitations, cannot scalably protect quantum information with a local decoding protocol.
Nevertheless, we still observe pseudo-threshold behavior, as discussed further below, and gain simple intuition for how LEC works using this model.
To extend our analysis to models that can, in principle, be protected by an LEC procedure, we consider the 2D (classical) Ising model, which realizes a classical memory~\cite{day2012ising}, and the 4D toric code, which realizes a quantum memory~\cite{Dennis_2002, doi:10.1142/S1230161210000023}.
Both have line-like syndrome excitations, which can be corrected via local operations~\cite{10.5555/3179532.3179533, toom1974, toom1980stable, Gray1999}.
With sufficiently low error rates, LEC circuits can protect their information indefinitely, making these codes \emph{self-correcting memories}.
Each model can be defined by its stabilizers, where their eigenvalues are either $+1$ or $-1$.
Then, each model encodes logical qubit(s) where all stabilizers have $+1$ eigenvalues.
Note that each lattice of the error-correcting code below has a periodic boundary condition.
\begin{itemize}
    \item 2D toric code~\cite{Kitaev.2003}:

    As in Fig. \ref{fig:1A-1_StabilizerDefinition}(a), data qubits live on edges of the 2D lattice, and plaquette-type ($\hat{A}_p$) and vertex-type ($\hat{B}_v$) stabilizers are tensor products of four Pauli $\hat{Z}$s (or $\hat{X}$s) on data qubits neighboring each plaquette $p$ (or vertex $v$).
    % :
    % \begin{equation}
    %     \hat{A}_p = \bigotimes_{e \in p} \hat{Z}_e, \quad \hat{B}_v = \bigotimes_{e \in v} \hat{X}_e
    % \end{equation}
    % where $e$ corresponds to the index of an edge neighboring the vertex $v$ or the plaquette $p$.
    %
    Two logical qubits are encoded with logical $\hat{Z}$ (or $\hat{X}$) that correspond to tensor products of $\hat{Z}$s (or $\hat{X}$s) along the topologically non-trivial loops.

    \item 2D Ising model~\cite{Dennis_2002}:

    As in Fig. \ref{fig:1A-1_StabilizerDefinition}(b), data qubits live on plaquettes of the 2D lattice, and vertical-type ($\hat{A}_v$) and horizontal-type ($\hat{A}_h$) stabilizers are tensor products of two $\hat{Z}$s on the data qubits adjacent to each vertical or horizontal edge.
    % \begin{align}
    %     \hat{A}_{v, (i,j)} &= \hat{Z}_{(i,j)} \otimes \hat{Z}_{(i,j+1)}, \\
    %     \hat{A}_{h,(i,j)} &= \hat{Z}_{(i,j)} \otimes \hat{Z}_{(i+1,j)}
    % \end{align}
    % where $(i,j)$ denotes a 2D coordinate of a plaquette.
    %
    Since the 2D Ising model can detect only a bit-flip error on the qubit, it encodes a classical memory.
    Also, one logical qubit is encoded that consists of all qubits with either $\ket{0}$ or $\ket{1}$.

    \item 4D toric code~\cite{Dennis_2002}:

    As in Fig. \ref{fig:1A-1_StabilizerDefinition}(c), data qubits live on the faces of the 4D lattice, and there are edge-type ($\hat{A}_e$) and cube-type ($\hat{B}_c$) stabilizers.
    Each stabilizer is a tensor product of six $\hat{Z}$s (or $\hat{X}$s) on data qubits neighboring each edge $e$ (or cube $c$).
    % \begin{equation}
    %     \hat{A}_e = \bigotimes_{f \in e} \hat{Z}_e, \quad \hat{B}_c = \bigotimes_{f \in c} \hat{X}_c,
    % \end{equation}
    % where $f$ corresponds to the index of a face neighboring the edge $e$ or the cube $c$.
    %
    Two logical qubits are encoded with logical $\hat{Z}$ (or $\hat{X}$) that correspond to tensor products of $\hat{Z}$s (or $\hat{X}$s) on the topologically non-trivial sheets.
\end{itemize}

% \bigskip\ptitle{Circuit-level syndrome extraction}

\begin{figure}[t]
    \centering
    \includegraphics[width = 0.48 \textwidth]{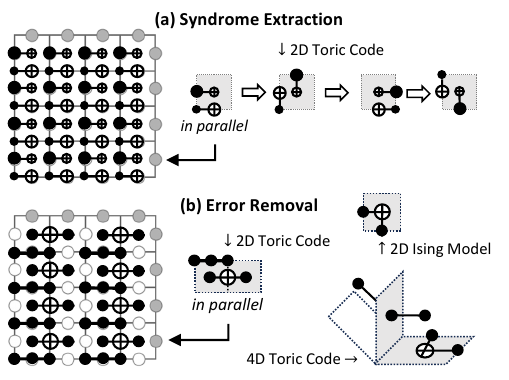}
    \caption{
    The gate-level implementation of syndrome extraction and error removal operation in each code.
    (a) Syndrome extraction operation using ancillas is a sequence of CNOT gates in parallel.
    An example of such a sequence with unit cells for 2D toric code is shown.
    Note that the CNOT gates are applied in the optimized order (see Appendix \ref{appendix: QECC-implementation}).
    % To store eigenvalues of stabilizers that consist of $\hat{Z}$s in ancillas, applied CNOT gates are controlled by data qubits and targeted on ancillas, and vice versa for stabilizers that consist of $\hat{X}$s, following the optimized order as discussed in Appendix \ref{appendix: QECC-implementation}.
    %
    (b) Error removal operation to reduce error density of data qubits controlled by ancillas is a single application of CCX or CCZ gates in parallel.
    An example of the simplest such operation with a unit cell for each code is shown.
    To remove $\hat{X}$ (or $\hat{Z}$) errors, CCX (or CCZ) gates are applied.
    }
    \label{fig:1A-2_CircuitOperations}
\end{figure}

The syndrome extraction circuits are used to measure stabilizers via ancillas.
For each of the above error-correcting codes, the stabilizers are a tensor product of $\hat{Z}$s or $\hat{X}$s on the data qubits.
Since $\hat{Z}$ and $\hat{X}$ anti-commute with each other, eigenvalue of stabilizer $\hat{S}_Z$ (or $\hat{S}_X$) is $(-1)^{n_s}$ where $n_s$ is a number of $\hat{X}$ (or $\hat{Z}$) errors on data qubits that support the stabilizer.
Due to the commutation relation between a CNOT gate and $\hat{X}$ and $\hat{Z}$, a sequence of CNOT gates performs a syndrome extraction.
For example, for 2D toric code, we put an ancilla initialized to $\ket{0}$ on every plaquette and then apply the sequence of parallelized CNOT gates as shown in Fig. \ref{fig:1A-2_CircuitOperations}(a).
Then, each ancilla remains as $\ket{0}$ if an eigenvalue of corresponding $\hat{S}_Z$ is +1 and flips into $\ket{1}$ otherwise.
%
% This provides a simple procedure to extract syndromes.
% %
% For 2D toric code, we consider an ancilla initialized to $\ket{0}$ for every plaquette.
% %
% Then, we apply a sequence of four CNOT gates controlled by data qubits at edges neighboring the plaquette and targeted on the ancilla.
% %
% Due to the commutation relation between a CNOT gate and $\hat{X}$ and $\hat{Z}$ operator, we obtain
% \begin{equation}
%     \hat{X}^{(1 + S_Z)/2} \ket{0} = \begin{cases}
%         \ket{0} & \text{if } S_Z = +1\\
%         \ket{1} & \text{if } S_Z = -1
%     \end{cases},
% \end{equation}
% where $S_Z$ is an eigenvalue of $\hat{S}_Z$.
%
% Similarly, using two Hadamard gates and a sequence of four CNOT gates controlled by the ancilla and target on data qubits, we obtain ancillas on vertices storing eigenvalues of $\hat{S}_X$.
% %
% For a more efficient syndrome extraction operation, we parallelize the CNOT gates to the entire 2D toric code lattice as shown in Fig. \ref{fig:1A-2_CircuitOperations}(a).
% %
Similar sequences of CNOT gates in parallel can detect errors by syndrome extraction for the 2D Ising model and 4D toric code.
Note that we assume a reset of ancillas to their initial states $\ket{0}$ before a new syndrome extraction operation.

% \bigskip\ptitle{Circuit-level error removal}

% \todo{Explain higher-level here.}
The error removal operations are more complex and involve applying correction gates conditioned on the state of the ancillas.
Let $\ket{A_1}$ and $\ket{A_2}$ be two ancillas that store an eigenvalue of stabilizer $\hat{S}_{Z,A_1}$ and $\hat{S}_{Z,A_2}$, respectively.
Then, applying CCX gate controlled by these two ancillas --- without knowing their quantum states --- and targeted on the data qubit $\ket{D}$ results in $\hat{X}\ket{D}$ if both of the controlled qubits are $\ket{1}$s and $\ket{D}$ otherwise.
% \begin{equation}
%     \ket{D} \to \begin{cases}
%          & \text{ if } \ket{A_1} = \ket{A_2} = \ket{1}\\
%         \ket{D} & \text{ otherwise}
%     \end{cases}
% \end{equation}
%
As the simplest scenario, consider these two stabilizers share one supporting data qubit $\ket{D}$.
% $\hat{S}_{Z,A_1}$ and $\hat{S}_{Z,A_2}$ share one supporting data qubit $\ket{D}$.
%
If both stabilizers have $-1$ eigenvalues, we have $\ket{A_1} = \ket{A_2} = \ket{1}$ by syndrome extraction and there are an odd number of errors on data qubits supporting each of these stabilizers.
%$S_{Z,A_1} = S_{Z,A_2} = -1$, i.e. $\ket{A_1} = \ket{A_2} = \ket{1}$, there are odd number of errors on data qubits supporting $\hat{S}_{Z,A_1}$ and in $\hat{S}_{Z,A_2}$ each.
%
When data qubits have low error density, it is likely that the shared data qubit $\ket{D}$ has an $\hat{X}$ error, which gets corrected by applying the above CCX gate.
We can similarly remove a $\hat{Z}$ error on the data qubit by applying a CCZ gate.
% We can similarly remove a $\hat{Z}$ error on the data qubit $\ket{D}$ by applying CCZ gate controlled by ancillas $\ket{A_3}$ and $\ket{A_4}$, storing an eigenvalue of stabilizer $\hat{S}_{X,A_3}$ and $\hat{S}_{X,A_4}$ respectively, and targeted on $\ket{D}$.
%
The simplest gate layout for an error removal operation in each code is shown in Fig. \ref{fig:1A-2_CircuitOperations}(b).

% \bigskip\ptitle{Comparison with \emph{global decoding} idea}

A typical \emph{global decoding} approach also uses ancilla-based syndrome extraction operations.
However, instead of performing the error removal operation via local three-qubit gates, all the ancillas are projectively measured after the syndrome extraction operation.
Then, by globally processing readout outcomes, a classical decoding algorithm infers the most likely underlying errors, which can be subsequently corrected by applying local Pauli operators.
However, in our LEC scheme, only certain local correction operations are implemented, leading to the existence of additional \emph{uncorrectable} error configurations.

% \bigskip\ptitle{Reference to relevant Appendix}

% We explain how we implement a code base for circuit-level error correction benchmark simulation for each error-correcting code in more detail in Appendix \ref{appendix: QECC-implementation}.

\subsection{Extended LEC operations} \label{sec: IIB}

\begin{figure}[t]
    \centering
    \includegraphics[width = 0.48 \textwidth]{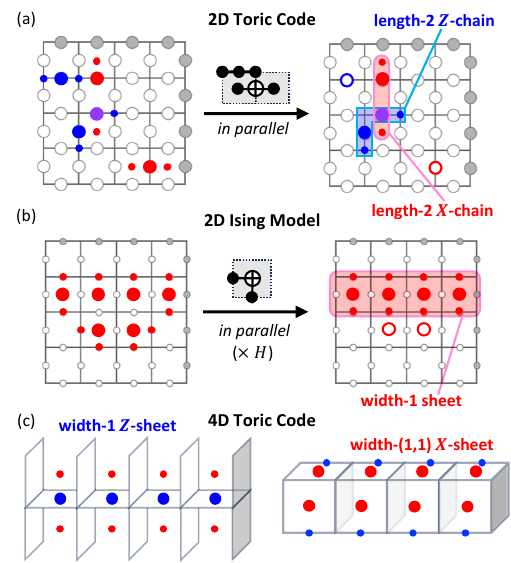}
    \caption{Uncorrectable error pattern by the simplest error removal operation in each code, as introduced in Fig. \ref{fig:1A-2_CircuitOperations}(b).
    (a) In 2D toric code, the given error removal operation removes only length-1 error chains that consist of a single $\hat{X}$ or $\hat{Z}$ error on a data qubit.
    However, error chains with more than one error are uncorrectable by the given operation.
    (b,c) In the 2D Ising model and 4D toric code, illustrated membrane-like error sheets bounded by parallel 1D syndrome lines that are periodic around the lattice are uncorrectable by the arbitrary repetition of Toom's rule error removal operation.
    }
    \label{fig:1B_Uncorrectable}
\end{figure}

% \ptitle{General error pattern in each code}

The syndrome extraction and error removal circuits form the building blocks of LEC circuits.
To understand the capabilities of LEC, we start by discussing how different kinds of errors behave under these circuits.
Each code has a specific error pattern where syndromes, which are $-1$ eigenvalues of stabilizers, form a boundary of errors on neighboring data qubits.
As shown in the left of Fig. \ref{fig:1B_Uncorrectable}(a), 2D toric code has 1D chains of errors on data qubits, which share a common adjacent plaquette or vertex, with endpoints of syndromes.
These error chains can be classified by their lengths, i.e. how many errors on data qubits are between two syndrome endpoints to compose the chains.
Here, we define the length of an error chain based on the shortest chain up to multiplication with stabilizers.
Applying the simplest error removal operation can remove length-1 $X$-chains or $Z$-chains that consist of a single $\hat{X}$ or $\hat{Z}$ error on a data qubit~\cite{Smith_2023}.

In contrast, as shown in the left of Fig. \ref{fig:1B_Uncorrectable}(b) and (c), the 2D Ising model and 4D toric code have 2D membranes of data qubit errors with 1D boundaries of syndromes.
For these codes, repeating Toom's rule operation is an established cellular automata decoding protocol~\cite{toom1974, toom1980stable}.
% Note that the introduced layout for the 2D Ising model and 4D toric code is called a Toom operation --- from Toom's rule that is a cellular automata decoding protocol for these codes~\cite{toom1974, toom1980stable, Gray1999}.
%
This operation can be implemented with CCX or CCZ gate as depicted in Fig. \ref{fig:1B_Uncorrectable}(b) followed by syndrome extraction.
As this operation effectively ``shrinks'' the size of a 2D error membrane, its repetition can remove all errors in an infinite-sized system if the error density of data qubits is below some threshold~\cite{Gray1999, kazemi2021genuine, liu2023dissipative}.
% and effectively ``shrinks'' the size of a 2D error membrane.
% %
% In the limit where a system size $L \to \infty$, the repetition of this local operation followed by syndrome extraction operation can remove all errors in the system if the error density of data qubits is below some threshold~\cite{Gray1999, kazemi2021genuine, liu2023dissipative}.
%
Due to this property, the 2D Ising model and 4D toric code are called \emph{self-correcting memories}.

% \bigskip\ptitle{Uncorrectable error pattern in each code}

However, we observe error patterns that are \emph{uncorrectable} by the simplest error removal operation as shown in Fig. \ref{fig:1B_Uncorrectable}.
%
% These error patterns limit the performance of the local decoding protocol for each code.
%
In 2D toric code, the simplest error removal operation cannot remove nor reduce the error chains with length $\ge 2$.
Also, for the finite-sized 2D Ising model and 4D toric code, some error configurations are uncorrectable by repeated Toom's rule operations --- even if the error density of data qubits is below the threshold.
%
% We illustrate the examples of such error patterns uncorrectable by repeated Toom's rule operation as in Fig. \ref{fig:1B_Uncorrectable}(b) and (c) for the system size $L = 4$.
%
These configurations, also called as \emph{energy barriers}~\cite{Brown_2016_QM, Breuckmann_2018, energy-barrier-1}, form periodic 2D sheets of data qubit errors surrounded by \emph{parallel} 1D loops of syndromes.
We call these error patterns error \emph{sheets} with width defined as the distance between the boundary 1D syndrome loops.
Therefore, each error-correcting code has an issue of error patterns uncorrectable by the simplest error removal operation arising from different reasons: the operation's limitation for 2D toric code and a system's finiteness for 2D Ising model and 4D toric code.

% \subsection{Extended LEC operations} \label{sec: IIC}

% \ptitle{General idea to resolve uncorrectability}

Now, we introduce how to resolve this issue of uncorrectable error patterns that limit the performance of the local decoding protocol.
We design a set of operations for each code such that any single low-weight error pattern can, in principle, be corrected.

\begin{figure}[t]
    \centering
    \includegraphics[width = 0.48 \textwidth]{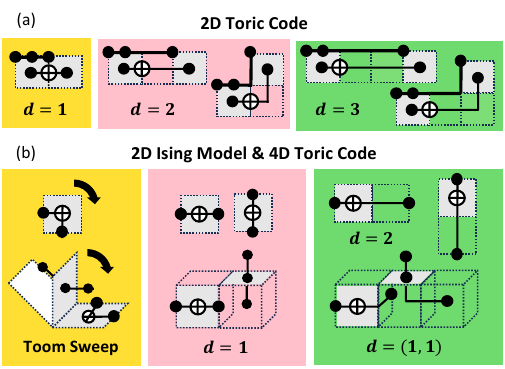}
    \caption{Unit cells of extended error removal operations for each code.
    To address the uncorrectable error patterns shown in Fig. \ref{fig:1B_Uncorrectable}, we hierarchically introduce new error removal operations.
    (a) In the 2D toric code, we include $d = N$ operations, for $N \le 3$, that can reduce length-$N$ chains into length-$(N-1)$ chains.
    (b) For the 2D Ising model and 4D toric code, we extend Toom's rule operation to \emph{Toom Sweep} operations as generated by 90$^\circ$ rotations of the original Toom's rule operation.
    Additionally, we introduce $d=N$ operations, for $N \le 2$ in the 2D Ising model and for $N = 1$ and $(1,1)$ in 4D toric code, that can reduce width-$N$ sheets into width-$(N-1)$ sheets.
    Note that $d=(1,1)$ operation for 4D toric code can reduce width-$(1,1)$ sheets into width-1 sheets.
    }
    \label{fig:1C_ExtendedOperations}
\end{figure}

% \bigskip\ptitle{Extension for 2D toric code}

For 2D toric code, we categorize the simplest removal operations as $d=1$ operations, because they can reduce length-1 error chains into length-0 error chains, i.e. no error.
% the extended set as illustrated in Fig. \ref{fig:1C_ExtendedOperations}(a) starts from the simplest $d=1$ operations.
% %
% Note that the $d=1$ operations can reduce length-1 error chains into length-0 error chains, i.e. no error.
%
As shown in Fig. \ref{fig:1C_ExtendedOperations}(a), we generalize this to $d=N$ operations that can reduce length-$N$ error chains into length-$(N-1)$ error chains.
We include up to $d \le 3$ operations in our optimization.

% \bigskip\ptitle{Extension for self-correcting memories}

In the 2D Ising model and 4D toric code, we extend the simplest operations in two ways as shown in Fig. \ref{fig:1C_ExtendedOperations}(b).
%the extended set as illustrated in Fig. \ref{fig:1C_ExtendedOperations}(b) starts from the simplest Toom operations.
%
First, we include the Toom Sweep operations generated by 90$^\circ$ rotations of the Toom operations, which were shown to reduce the occurrence of uncorrectable membrane-like error patterns~\cite{Kubica_2019, Vasmer_2021}.
Moreover, to enable the removal of such error patterns, we additionally introduce operations designed to remove errors between the two parallel 1D boundaries of syndromes.
These operations take the same form as the $d = N$ operations for 2D toric code and can reduce width-$N$ error sheets to have narrower widths.
We include two lowest-weight operations for both codes --- $d=1$ and $d=2$ operations for the 2D Ising model and $d=1$ and $d=(1,1)$ operations for the 4D toric code.

% \bigskip\ptitle{Remaining issue: how to optimize?}

If we did read out the syndromes, then we would apply the proper combination of these operations within the extended set to correct any error chain with length $\le 3$ or any error sheet with width $\le 2$ and $\le (1,1)$.
However, because the correction steps are applied via noisy controlled gates without the explicit knowledge of syndromes, constructing high-performance sequences of operations is quite challenging.
To overcome this challenge, we develop an RL framework to optimize LEC circuits, which are sequences of the error removal operations as well as syndrome extraction operations.
 
% \bigskip
% \vspace{2em}

\section{RL Optimization Framework} \label{sec: III}

% \ptitle{Section overview}

In this section, we provide details about the RL framework to optimize LEC circuits.
Section \ref{sec: IIIA} introduces the setting of the RL framework in terms of its agent, environment, and the interplay between them.
Then, Section \ref{sec: IIIB} discusses the result of depth and layout optimization of LEC circuits by the RL framework with varying multi-qubit gate fidelities.

RL is a machine learning paradigm where an \emph{agent} learns which \emph{actions} to play to maximize the \emph{reward} through its interplay with an \emph{environment}~\cite{10.5555/1622737.1622748,foster2023foundationsreinforcementlearninginteractive}.
In our setting, the LEC circuit builder (or agent) learns which sequence of LEC gate operations (or actions) maximizes the decoding success rate (or reward) through its interplay with the circuit-level simulator (or environment) until the convergence of the agent-built LEC circuit.
Due to the absence of mid-circuit readout in our scheme, our environment does not provide the \emph{observation} to the agent, unlike traditional RL setting~\cite{10.5555/1622737.1622748, foster2023foundationsreinforcementlearninginteractive}.

We aim to optimize an ordered sequence of discrete actions where an optimizer only observes the reward of the full sequence.
Since the sequence is not adjusted during the execution and the reward structure is ``opaque'' to the optimizer, our setting can be classified as an \emph{open-loop control} or \emph{black-box sequential optimization}.
Our problem features (1) a large search space of size between $10^{55}$ and $10^{85}$ possible ordered sequences, (2) the importance of discrete actions' order in the sequence, and (3) costly reward computation --- especially for 4D toric code.

% \bigskip\ptitle{Interplay between agent \& env}

Due to these features, we choose a proximal policy optimization (PPO) among RL techniques as our optimizer~\cite{schulman2017proximal}.
The RL algorithms have been used to robustly and sample-efficiently optimize the sequence of actions in large discrete action spaces~\cite{dulacarnold2016deepreinforcementlearninglarge,pinto2017robustadversarialreinforcementlearning,popov2017dataefficientdeepreinforcementlearning,tuan2018proximal}.
In particular, such algorithms have been successfully applied to the open-loop control problems~\cite{NIPS1996_ab1a4d0d, PhysRevFluids.6.053902, raffin2024openloopbaselinereinforcementlearning}.
More recently, the PPO algorithm has been used to optimize the quantum gate sequence, such as preparing multi-qubit GHZ states~\cite{kuo2021quantum} or fault-tolerant logical states in small quantum error-correcting codes~\cite{zen2024quantum}.
Although there are alternative optimizing strategies including variations of genetic algorithms~\cite{Katoch2021} or simulated annealing~\cite{LIU2021310}, we find that the RL method is particularly suitable considering the purpose and features of our problem.
% Additionally, we highlight that our RL framework avoids two common limitations of RL optimizers: (1) training cost and (2) training instability.
%
% Our RL framework has a relatively small training cost compared to the usual recurrent neural network (RNN) or transformer due to a smaller neural network being used, and it shows a consistent performance over multiple executions of training (see Appendix \ref{appendix: RL-stability}).

\begin{figure}[b]
    \centering
    \includegraphics[width = 0.48 \textwidth]{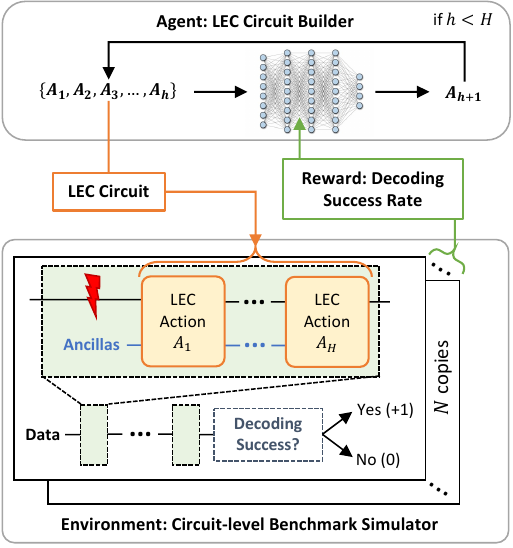}
    \caption{
    Overview of our reinforcement learning (RL) framework that consists of an agent, environment, and their interactions to optimize the LEC circuit for a given code and error model.
    By choosing the $(h+1)$-th action based on all previous actions, the agent's policy network builds an LEC circuit, which is a sequence of fixed circuit depth $H$ number of LEC actions, as the environment's input.
    Then, the environment benchmarks a decoding success rate given this LEC circuit with a fixed error model by simultaneously evaluating the success or failure of decoding for $N$ copies of codes after multiple LEC cycles, which are applications of LEC circuits after the introduction of new errors.
    This rate is then used as a reward to update the agent's neural network parameters.
    These steps are repeated until the perceived reward saturates, i.e. until the agent receives the same reward signal and does not design a new LEC circuit for multiple network updates.
    } 
    \label{fig:2A_SummaryRL}
\end{figure}

\subsection{Setting of RL framework}\label{sec: IIIA}

We summarize the specific interplay between the RL agent and environment as in Fig. \ref{fig:2A_SummaryRL} and Algorithm \ref{alg: RL}.
\begin{algorithm}
\caption{RL training algorithm}\label{alg: RL}
\KwData{Circuit depth $H$; action space $\mathcal{A}$; code type and parameters; error model parameters}
\KwResult{RL-optimized LEC circuit}

LEC circuit $\gets$ EmptyList\;
Initialize parameter $\vec{\theta}$ of neural network $f^{\vec{\theta}}$\; 
\Repeat{LEC circuit stays the same for multiple updates}{
    LEC circuit = BuildLEC($f^{\vec{\theta}}$, $H$, $\mathcal{A}$)\;
    Reward = ComputeReward(LEC circuit, code type and parameters, error model parameters)\;
    $\vec{\theta}$ := UpdateNetworkParameters($f^{\vec{\theta}}$, Reward);
    }
\end{algorithm}

First, the RL agent produces an LEC circuit based on its internal parameters.
Then, the RL environment takes this LEC circuit as an input to compute its decoding success rate with fixed code and error model parameters.
Such a rate is used as a reward to update the internal parameters of the RL agent.
This training process is repeated to maximize this reward until the resulting LEC circuit converges.

Let's further discuss each component of our RL framework: an agent, environment, and algorithm updating the agent's neural network based on reward from the environment:
\begin{itemize}
    \item The agent has a neural network $f^{\vec{\theta}}$ with vectorized internal parameters $\vec{\theta}$ that takes a sequence of $h$ (for $h < H$) actions as input and returns the next action as output, i.e. 
    \begin{equation}
        f^{\vec{\theta}}: \mathcal{A}^{\otimes h} \to \mathcal{A},
    \end{equation}
    where $\mathcal{A}$ denotes a set of all action candidates that depend on a choice of the error-correcting code.
    Repeating such next-action decision-making for $H$ times results in building the LEC circuit with the circuit depth $H$.
    See Appendix \ref{appendix: RL-agent} for more details.

    \item The environment computes the decoding success rate by a circuit-level Monte Carlo simulation of performing multiple rounds of applying an LEC circuit produced by the RL agent.
    This simulation takes a code type, error model parameters, and LEC circuit as inputs.
    Here, we consider two different error sources: (1) \emph{ambient} errors introduced every LEC round before applying the LEC circuit with the rate $p_{\rm{amb}}$ and (2) \emph{gate} errors introduced after applying each parallelized multi-qubit gate with rate $p_{\rm{gate}}$.
    Note that the reward computation is based on the \emph{average} performance of an LEC circuit produced by the RL agent, because the decoder cannot access syndrome information in our measurement-free scheme.
    See Appendix \ref{appendix: RL-env} for more details.

    \item We utilize the Python-based package Stable-Baselines3 to perform the training to implement the RL framework~\cite{stable-baselines3}.
    Recall that we update the agent's network parameters from the reward with the PPO algorithm.
    To obtain the best-performing LEC circuit, we run the RL training four times independently until the training reaches its termination condition.
    Among the final LEC circuits from four trained models, we choose the one that maximizes the reward.
    % For updating the agent's network parameters based on the reward from the environment, we adopt a proximal policy optimization (PPO) algorithm~\cite{schulman2017proximal}.
    %
    % This algorithm has been widely used for optimizing a sequence of actions to perform a difficult task~\cite{tuan2018proximal, kuo2021quantum}.
\end{itemize}

Appendix \ref{appendix: RL} includes more details on our RL framework: a training termination condition (see Appendix \ref{appendix: stop-training}), hyperparameters (see Appendix \ref{appendix: RL-hyperparameters}), and stability of training (see Appendix \ref{appendix: RL-stability}).

\subsection{Optimizing depth and layout of LEC circuit} \label{sec: IIIB}

% \ptitle{Choice of code \& error model parameters}

%% Training Info
In each error-correcting code, we provide varying gate error rates $p_{\rm{gate}}$ as an input to our RL framework to optimize the length and layout of LEC circuits for each input $p_{\rm{gate}}$.
We choose the scale of $p_{\rm{gate}}$ to be $\sim 10^{-4}$ for 2D toric code, $\sim 10^{-3}$ for 2D Ising model, and $\sim 10^{-5}$ for 4D toric code (see Table \ref{tab: parameters} in Appendix \ref{appendix: RL-env} for other code and error model parameters).
%
% Then, we fix the other parameters for the code and the error model as shown in Table \ref{tab: parameters} to perform RL training.

\begin{figure}[t]
    \centering
    \includegraphics[width = 0.48 \textwidth]{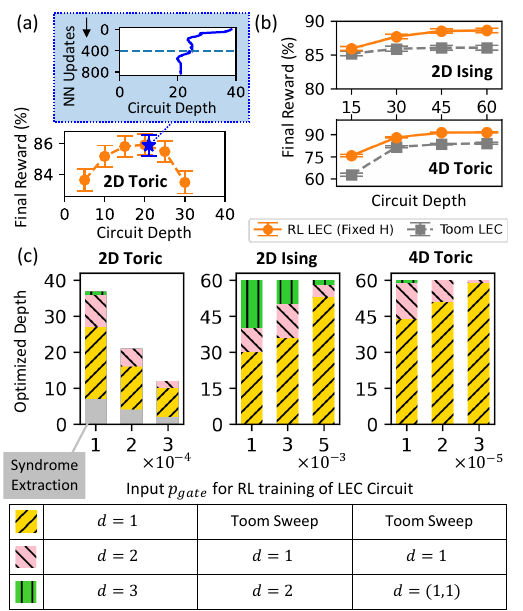}
    \caption{Optimization of depth and layout of the LEC circuit by the RL framework in each error-correcting code.
    %
    % We use code and error model parameters as in Table \ref{tab: parameters}.
    %
    Each final reward is computed with 10000 independent samples (see Appendix \ref{appendix: error bar} for the error bar).
    %
    % Our computation for the error bar of the final reward is discussed in Appendix \ref{appendix: error bar}.
    %
    (a) For 2D toric code, we show the RL training with variable depth with training $p_{\rm{gate}} = 2 \times 10^{-4}$: as number of neural network (NN) policy updates increases, the LEC circuit's variable depth converges.
    By comparing the final reward of the resulting LEC circuit trained by the RL framework with fixed vs. variable depth setting, we confirm that the obtained circuit depth is effectively optimized.
    (b) For the 2D Ising model and 4D toric code, we show the final reward after RL training vs. fixed circuit depth with training $p_{\rm{gate}} = 3\times 10^{-3}$ for 2D Ising model and $2\times 10^{-5}$ for 4D toric code.
    (c) For each training $p_{\rm{gate}}$, we classify the components of RL-optimized LEC circuits as categorized in Fig. \ref{fig:1C_ExtendedOperations}.
    }
    \label{fig:2B_OptimizedDepthLayout}
\end{figure}

% \bigskip\ptitle{RL framework with fixed vs. variable $H$}

From RL training with varying circuit depth $H$, we observe that there exists an \emph{optimal} circuit depth for 2D toric code as shown in Fig. \ref{fig:2B_OptimizedDepthLayout}(a).
%
% Thus, we construct the RL framework that optimizes both circuit depth and layout (see Appendix \ref{appendix: RL-agent} for details).
%
As shown in Fig. \ref{fig:2B_OptimizedDepthLayout}(c), we observe that the optimized circuit depth $H_{\rm{opt}}$ for 2D toric code depends on training $p_{\rm{gate}}$: $H_{\rm{opt}}$ decreases as training $p_{\rm{gate}}$ increases.
This trend is because faulty LEC action causes a trade-off between error removal by parallel CCX or CCZ gate operation and error introduced by a gate error model.
Even in $p_{\rm{gate}} = 0$ case, error removal by $h$-th LEC action $A_h$ gets marginalized as $h$ increases.
When $p_{\rm{gate}}$ increases, $A_h$ for large $h$ introduces more errors than it removes.
We further observe that higher-weight LEC actions ($d > 1$) are more frequently used for smaller training $p_{\rm{gate}}$.
This trend is because higher-weight error chains are less common than lower-weight error chains in the lattice and thus lower-weight LEC actions benefit more from the tradeoff of faulty LEC actions for smaller training $p_{\rm{gate}}$.

% \bigskip\ptitle{Detail on 2D Ising model \& 4D toric code}

Unlike in 2D toric code, $H > H_{\rm{c}} \approx 45$ shows a saturation of final reward in the self-correcting memories as shown in Fig. \ref{fig:2B_OptimizedDepthLayout}(b).
This is consistent with the system entering a steady state, where an action coming after some $H_{\rm{c}}$ achieves a balance between errors that it newly introduces by finite gate error or removes.
Note that as $H \to \infty$, we expect the reward to slowly go to zero due to the rare possibility of high-weight uncorrectable errors.
Thus, for the 2D Ising model and 4D toric code, we use the RL framework that optimizes circuit layout with fixed $H = 60$ (see Appendix \ref{appendix: RL-agent} for details).
However, just as in 2D toric code, Fig. \ref{fig:2B_OptimizedDepthLayout}(c) also shows that new LEC actions ($d = N$) are more frequently used for smaller training $p_{\rm{gate}}$ with the same reasoning.
Moreover, in both the 2D Ising model and 4D toric code, these new LEC actions appear only after a repetition of multiple Toom Sweep actions.
For all training $p_{\rm{gate}}$, RL-optimized LEC circuits start with $\ge 10$ Toom Sweep actions for 2D Ising model and $\ge 19$ Toom Sweep actions for 4D toric code.
This is because the membrane-like uncorrectable error configurations illustrated in Fig. \ref{fig:1B_Uncorrectable} appear only after applying an iteration of Toom's rule decoding.

\section{Memory Lifetime Extension by LEC} \label{sec: IV}

% \ptitle{Section overview}

In this section, we provide a thorough analysis of the memory lifetime extension by the LEC circuits and the comparison between RL-optimized LEC circuits (or \emph{RL LEC}) vs. conventional LEC circuits (or \emph{conventional LEC}) in each code.
Recall that the conventional LEC for 2D toric code is called a nearest-neighbor LEC circuit (or \emph{NN LEC})~\cite{Smith_2023} and that for 2D Ising model and 4D toric code is called a Toom's rule-based LEC circuit (or \emph{Toom LEC})~\cite{10.5555/3179532.3179533}.
%
% Section \ref{sec: IVA} analyzes the suppression of uncorrectable errors by RL LEC in comparison with conventional LEC.
%
Section \ref{sec: IVB} explains the idea of effective code distance that characterizes the lifetime of a memory encoded in the error-correcting code.
Then, Section \ref{sec: IVC} provides a comparison of this parameter for RL vs. conventional LEC based on the benchmark simulation results.
To explain why RL LEC performs better than conventional LEC, we provide the error statistics analysis in Appendix \ref{appendix: classification-detail}.

% \subsection{Effect of LEC circuits on error statistics}\label{sec: IVA}

\subsection{Idea of effective code distance} \label{sec: IVB}

We define the \emph{average memory lifetime} $T$ encoded in each code to be an average number of LEC rounds before the final recovery step calls decoding failure for the first time.
%
% Due to its relation with logical error rate, the lifetime of memory is extended more effectively by RL LECs than conventional ones.
%
We now introduce a parameter that represents the faulty LEC circuit's performance in extending the memory lifetime --- called \emph{effective code distance}.

% \ptitle{Memory lifetime vs. $p_{\rm{amb}}$ in No LEC}

First, let's consider a case without LEC, i.e. error with $p_{\rm{amb}}$ introduced per qubit per round until the final recovery failure after multiple rounds is evaluated by performing a perfect global decoding.
In this case, the error rate is accumulated approximately $p_{\rm{amb}}$ every round, and the average memory lifetime $T$ can be approximated as
\begin{equation} \label{eq:T_noLEC}
    T \sim \frac{p_{\rm{th}}}{p_{\rm{amb}}},
\end{equation}
where $p_{\rm{th}}$ refers to a threshold error rate for the final decoding success evaluation step.
Note that $p_{\rm{th}}$ depends on a type of code: 2D toric code has $p_{\rm{th}} \approx 12\%$, 2D Ising model has $p_{\rm{th}} = 50$\%, and 4D toric code has $p_{\rm{th}} \approx 4.5\%$.
This behavior is because, for a large enough system size, the final recovery step can be approximated to perform successful decoding if the system's error rate is below $p_{\rm{th}}$ and failed decoding otherwise.

Now, let's consider a case with LEC: we repeat error generation with rate $p_{\rm{amb}}$ followed by an LEC circuit application until the decoding failure is evaluated.
First, suppose that gates are perfect, i.e., $p_{\rm{gate}} = 0$.
Since an LEC circuit reduces error density in the system, the LEC circuit is expected to suppress the accumulated rate of errors $p_{\rm{acc}}$ in each LEC round:
\begin{equation} \label{eq:P_fail_LEC}
    p_{\rm{acc}} \propto (p_{\rm{amb}})^{D_{\rm{eff}}},
\end{equation}
where the power law exponent is called \emph{effective code distance} $D_{\rm{eff}}$ of the LEC circuit.
Combining Equation \ref{eq:T_noLEC} and \ref{eq:P_fail_LEC}, we obtain that
\begin{equation}
    T \sim \frac{p_{\rm{th}}}{p_{\rm{acc}}} \sim (p_{\rm{amb}})^{-D_{\rm{eff}}}.
\end{equation}
Thus, for the same final recovery step and $p_{\rm{amb}}$ each LEC round, $T$ gets extended as $D_{\rm{eff}}$ increases.
Since different LEC circuits suppress error accumulation by different amounts, $D_{\rm{eff}}$ characterizes the performance of the LEC circuit in extending the memory lifetime.

% \bigskip\ptitle{Physical meaning of $D_{\rm{eff}}$}

In particular, $D_{\rm{eff}}$ depends on the size of the \emph{uncorrectable error configuration} by the given LEC circuit.
Without LEC, every error in the system survives, gets passed to the next round, and contributes to the accumulation of error density.
However, with LEC, primarily the ``uncorrectable'' error configuration by the entire LEC circuit survives and contributes to the error density accumulation.
We summarize the scale of uncorrectable error by LEC compared to logical error in Table \ref{tab:scale-uncorrectable}.

\begin{table}[h]
\renewcommand{\arraystretch}{1.5}
\begin{tabular}{ |c|c|c| } 
 \hline
  & LEC's uncorrectable error & Logical error \\ 
 \hline
 2D toric code & $D$ & $L$ \\ 
 \hline
 2D Ising model & $D \times L$ & $L^2/2$ \\ 
 \hline
 4D toric code & $D \times L$ & $L^2$ \\
 \hline
\end{tabular}
\caption{Number of data qubit errors to compose an effectively uncorrectable error pattern by entire LEC circuit vs. a logical error in each code.
$L$ denotes a system size, and $D$ denotes some constant that depends on the LEC circuit.}
\label{tab:scale-uncorrectable}
\end{table}
\renewcommand{\arraystretch}{1}

% \bigskip\ptitle{Effect of faulty gates}

\begin{figure}[t]
    \centering
    \includegraphics[width = 0.48 \textwidth]{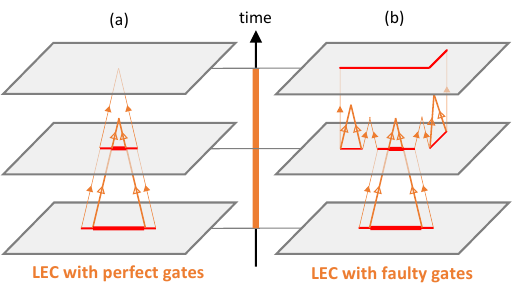}
    \caption{Effect of faulty gates on effective code distance by comparing a correction of lower-weight (thick red line) vs. higher-weight (thin red line) initial error configurations.
    Local decoding with the same LEC circuit over time is schematically shown for perfect vs. faulty gates.
    (a) When gates are perfect, the LEC circuit removes both of the initial error configurations at the end.
    (b) When gates are faulty, the LEC circuit still removes an error configuration with a lower weight at the end by the local decoding process (thick orange lines with empty arrows).
    However, an error configuration with a higher weight ``grows'' over time by merging with gate errors introduced during the local decoding process (thin orange lines with filled arrows).
    Thus, we expect the effective code distance for a given LEC circuit to decrease as the circuit's gate error rate increases.
    } 
    \label{fig:3B_FaultyDeff}
\end{figure}

% \bigskip

%% Faulty LEC (Fig. 8)
Although the above discussion is for LEC circuits with perfect gates, we also define $D_{\rm{eff}}$ for faulty LEC circuits as well (i.e., $p_{\rm{gate}} \neq 0$).
Note that the LEC circuit can be interpreted as a dynamical process.
The initial error configuration is newly introduced to the system by ambient or gate errors.
Then, the LEC circuit is applied to correct such an error pattern.
When gates are perfect, the decoding is performed only to these initial errors as shown in Fig. \ref{fig:3B_FaultyDeff}(a).
However, when gates are faulty, new errors introduced after each parallel gate operation can cause the growth of initial error configuration during the decoding process as shown in Figure \ref{fig:3B_FaultyDeff}(b).
Due to this effect, $D_{\rm{eff}}$ gets decreased when LEC circuits consist of faulty gates.
Also, these faulty gates impact the failure rate in each round as well due to this insertion of errors by gate errors during the local decoding process.
Thus, we model the average memory lifetime $T$ for a given LEC circuit with faulty gates as
\begin{equation} \label{eq:lifetime_model}
    T \approx C \left( \frac{1}{\alpha \times p_{\rm{amb}}} \right)^{D_{\rm{eff}}}.
\end{equation}

Both $D_{\rm{eff}}$ and $\alpha$ depend on the choice of error-correcting code, $p_{\rm{gate}}$, and an LEC circuit.
Physically, the parameter $D_{\rm{eff}}$ describes how fast such errors \emph{grow} over time, whereas the parameter $\alpha$ describes how many new errors (including gate errors) \emph{get introduced} every LEC round.
From Table \ref{tab:scale-uncorrectable}, we expect that (1) $D_{\rm{eff}}$ is independent of $L$ for 2D toric code and (2) $D_{\rm{eff}} \propto L$ for 2D Ising model and 4D toric code.
%
% This intuition is confirmed in the following Section \ref{sec: IVC}.
%
% Also, we discuss the detailed intuition for $\alpha$ in Appendix \ref{appendix:  other-lifetime-parameters}.
%
Our characterization of memory lifetime by a local decoder is different from the previous literature~\cite{10.5555/3179532.3179533}, which follows the quadratic relation between lifetime and ambient error rate.
This is due to the difference between gate error regimes of interest.

\subsection{Fitting effective code distance} \label{sec: IVC}

% \ptitle{Fitting $T$ vs. $p_{\rm{amb}}$}

\begin{figure}[t]
    \centering
    \includegraphics[width = 0.48 \textwidth]{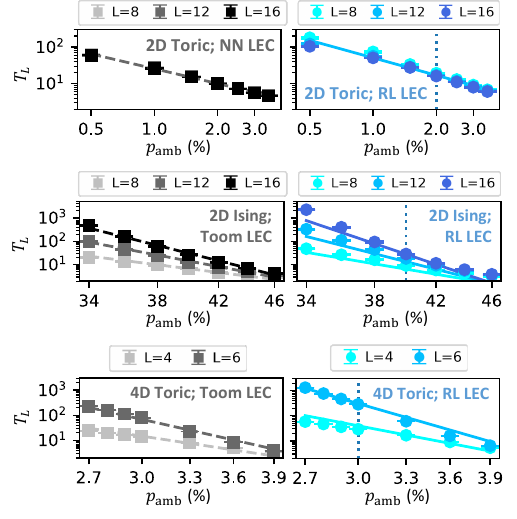}
    \caption{Fitting from log-log plot of average memory lifetime $T_L$ for each system size $L$ vs. ambient error rate $p_{\rm{amb}}$ for fixed benchmark gate error rate $p_{\rm{gate}}$ to compute effective code distance.
    In each code, we perform a simulation of multiple LEC cycles on 1000 independent samples to benchmark $T_L$, which is the number of LEC cycles until the final perfect decoding results in a logical failure (see Appendix \ref{appendix: error bar} for the error bar).
    %
    % Our computation for the error bar in average $T_L$ is discussed in Appendix \ref{appendix: error bar}.
    %
    We compare RL-optimized LEC circuit (blue gradation for varying $L$) vs. nearest-neighbor LEC circuit (NN LEC) in 2D toric code or Toom's rule-based LEC circuit (Toom LEC) in 2D Ising model and in 4D toric code (grey gradation for varying $L$).
    We train the RL model with $p_{\rm{gate}} = 1\times 10^{-3}, \text{ }1\times 10^{-4}, \text{ and }1\times 10^{-5}$ for 2D Ising model, 2D and 4D toric code, respectively.
    Note that a blue dashed line (:) corresponds to training $p_{\rm{amb}}$ for each code.
    } 
    \label{fig:3C-1_FittingDeff}
\end{figure}

\begin{figure*}[t]
    \centering
    \includegraphics[width = \textwidth]{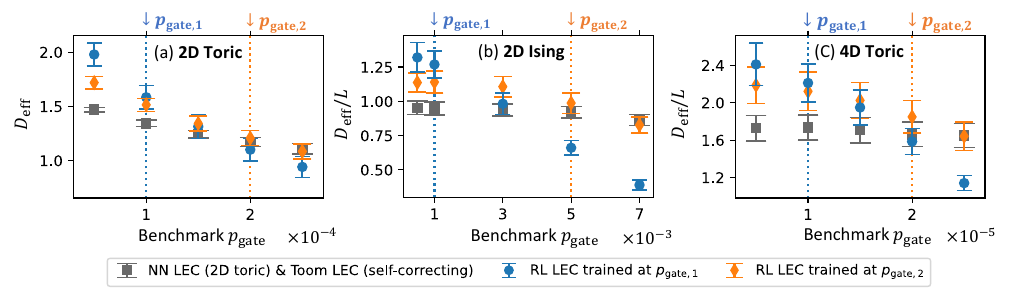}
    \caption{Fitted effective code distance $D_{\rm{eff}}$ for given LEC circuits vs. benchmark gate error rate $p_{\rm{gate}}$ in each code (see Appendix \ref{appendix: error bar} for the error bar).
    %
    % Note that each curve with the same color} represents $D_{\rm{eff}}$ vs. benchmark $p_{\rm{gate}}$ for the same LEC circuit.
    %
    % Our computation for the error bar of $D_{\rm{eff}}$ is discussed in Appendix \ref{appendix: error bar}.
    %
    Grey squares represent data with (a) nearest-neighbor LEC circuit (NN LEC) in 2D toric code and (b,c) Toom's rule-based LEC circuit (Toom LEC) in 2D Ising model and 4D toric code.
    Blue circles (or orange diamond) represent data with RL-optimized LEC circuit trained at $p_{\rm{gate}}$ with value (a) $1\times 10^{-4}$ (or $2\times 10^{-4}$) in 2D toric code, (b) $1\times 10^{-3}$ (or $5\times 10^{-3}$) in 2D Ising model, and (c) $1\times 10^{-5}$ (or $2\times 10^{-5}$) in 4D toric code.
    %
    % Orange diamonds represent data with RL-optimized LEC circuit trained at (a) $p_{\rm{gate}} = 2 \times 10^{-4}$ in 2D toric code, (b) $p_{\rm{gate}} = 5\times 10^{-3}$ in 2D Ising model, and (c) $p_{\rm{gate}} = 2\times 10^{-5}$ in 4D toric code.
    }
    \label{fig:3C-2_ComparingDeff}
\end{figure*}

%% Benchmark detail
In Equation \ref{eq:lifetime_model}, the parameter $C$, $\alpha$, and $D_{\rm{eff}}$ can be determined by fitting the data between average memory lifetime $T_L$ vs. ambient error rate $p_{\rm{amb}}$ as shown in Fig. \ref{fig:3C-1_FittingDeff}.
To achieve this fitting, we benchmark the average memory lifetime $T_L$ in each error-correcting code with varying input parameters.
Then, we compare the RL LEC trained with different $p_{\rm{gate}}$ and conventional LEC.
%
% In the 2D toric code, we compare the NN LEC and RL LEC trained with different $p_{\rm{gate}}$.
%
% In the 2D Ising model and 4D toric code, we compare the Toom LEC and RL LEC trained with different $p_{\rm{gate}}$.
%
% $T_L$ for each of these LEC circuits is benchmarked with $L \in \{8,12,16\}$ for 2D toric code and 2D Ising model and with $L \in \{4,6\}$ for 4D toric code.

%% Intro to fitting model
% Based on the data obtained from these benchmark simulations, we perform a fitting to find $D_{\rm{eff}}$ in each error-correcting code.
%
% Note that the fitting lines in Fig. \ref{fig:3C-1_FittingDeff} are obtained from the following fitting models.

In 2D toric code, $T$ is approximately independent of $L$ (see Appendix \ref{appendix: T-L-2Dtoric} for a caveat).
%
%as confirmed in Fig. \ref{fig:3C-1_FittingDeff} --- with a caveat as discussed in Appendix \ref{appendix: T-L-2Dtoric}.
%
Thus, from Equation \ref{eq:lifetime_model}, we fit the following model between $T$ and $p_{\rm{amb}}$:
\begin{equation} \label{eq: model-2Dtoric}
    \log_{10} (T) = - \textcolor{red}{D_{\rm{eff}}} \times \log_{10} (p_{\rm{amb}}) + \textcolor{red}{k},
\end{equation}
where $k=\log_{10}C - D_{\rm{eff}} \times \log_{10}\alpha$.
In 2D Ising model and 4D toric code, $D_{\rm{eff}}$ is expected to be proportional to $L$.
Thus, from Equation \ref{eq:lifetime_model}, we fit the following model between $T_L$ (average memory lifetime obtained by simulating a code with system size $L$) and $p_{\rm{amb}}$:
\begin{equation} \label{eq: model-self}
    \log_{10} (T_L) = - \textcolor{red}{\frac{D_{\rm{eff}}}{L}} \times L \times (\log_{10} (p_{\rm{amb}}) + \textcolor{red}{k_1}) + \textcolor{red}{k_2},
\end{equation}
where $k_1$ and $k_2$ are $\log_{10}\alpha$ and $\log_{10}C$, respectively.

With a fixed LEC circuit, $L$, and benchmark $p_{\rm{gate}}$, let each dataset be the pairs of benchmark $p_{\rm{amb}}$ and average lifetime.
Then, to perform a global fitting based on Equation \ref{eq: model-2Dtoric} and \ref{eq: model-self}, we use the symfit python package~\cite{Roelfs2017}.
Parameters $(D_{\rm{eff}}, k)$ for the 2D toric code and $(D_{\rm{eff}}, k_1, k_2)$ for the 2D Ising model and 4D toric code were shared by datasets varying $L$ with the same LEC circuit and benchmark $p_{\rm{gate}}$.
%
% Although we are mainly interested in $D_{\rm{eff}}}$ in each code, we also discuss $k_1$ fitting result based on our intuition of $\alpha$ in Appendix \ref{appendix: other-lifetime-parameters}.

%% p_gate vs. D_eff
After the fitting, we plot $D_{\rm{eff}}$ of each LEC circuit vs. benchmark $p_{\rm{gate}}$ as shown in Fig. \ref{fig:3C-2_ComparingDeff} (see Appendix \ref{appendix: other-lifetime-parameters} for $k_1$ fitting).
Let $p_{\rm{gate,c}}$ for RL LEC be a critical gate error rate such that RL LEC has higher $D_{\rm{eff}}$ compared to NN LEC (if 2D toric code) or Toom LEC (if 2D Ising model or 4D toric code).
In every error-correcting code, we observe that the RL LEC trained with lower $p_{\rm{gate}}$ shows higher fitted $D_{\rm{eff}}$ in small $p_{\rm{gate}}$ regime but smaller $p_{\rm{c}}$ compared to the LEC circuit trained with higher $p_{\rm{gate}}$.
For the LEC circuits that are trained with small $p_{\rm{gate}}$ as shown in Fig. \ref{fig:3C-2_ComparingDeff}, we additionally find that $p_{\rm{gate,c}}$ is larger than the training $p_{\rm{gate}}$.
Note that this property does not apply to LEC circuits trained with large $p_{\rm{gate}}$ (see Appendix \ref{appendix: large-gate-error}).
Also, in the 2D Ising model, the LEC circuit only with Toom actions shows fitted $D_{\rm{eff}}/L \approx 1.0$.
This is because the minimum uncorrectable error configuration by the Toom LEC circuit consists of $L$ number of data qubit errors as shown in Fig. \ref{fig:1B_Uncorrectable}(b).
Note that this observation and its explanation are consistent with the intuition from Table \ref{tab:scale-uncorrectable}.

\section{Application of LEC Scheme} \label{sec: V}

% \ptitle{Section overview}

% In this section, we first discuss the experimental implementation of the LEC scheme in Section \ref{sec: VA}.
%
In this section, we provide two specific examples of how this optimized LEC scheme could be used in various quantum information processing tasks.
Section \ref{sec: VB} explores how RL-optimized LEC circuits can be used to reduce the number of mid-circuit readouts required to preserve a quantum memory in 2D toric code.
On the other hand, Section \ref{sec: VC} discusses the dissipative preparation of topological phases with a given Hamiltonian.

\subsection{Reducing rate of mid-circuit readouts} \label{sec: VB}

% \ptitle{Goal: preserving memory under $P_{\rm{L}}$}

Section \ref{sec: IV} shows that LEC circuits can be used to remove errors from encoded qubits and to extend the memory lifetime.
Then, it is reasonable to ask whether such LEC circuits can be beneficial to extend memory lifetime in practice.
We focus on a 2D toric code to investigate the practical application.
In 2D toric code, using only LEC has limitations for a practical application, because memory lifetime extended by LEC does not scale with $L$ as observed in Fig. \ref{fig:3C-1_FittingDeff}.
Also, undecoded error chains get accumulated by the repetition of LECs, which implies that extending memory lifetime has some upper limit.
However, we can envision a decoding strategy combining the advantages of both LEC and global decoding.
In particular, repeated application of LEC can be used to reduce the rate of error accumulation of an encoded logical qubit.
Then, we can periodically perform the global decoding.
The key advantage of this approach is that LEC circuits can be applied more quickly compared to the global decoding that involves mid-circuit readouts.

\begin{figure}[t]
    \centering
    \includegraphics[width = 0.48 \textwidth]{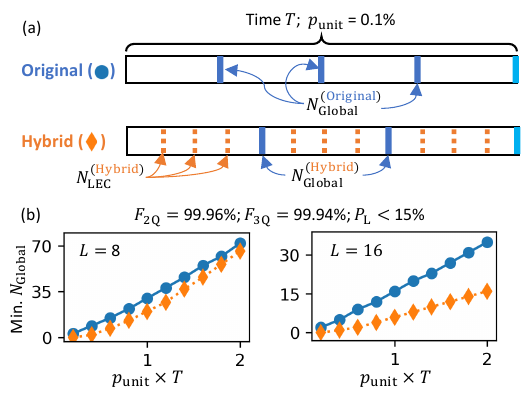}
    \caption{Original vs. Hybrid decoding protocol to preserve 2D toric code memory.
    (a) At each unit time, new errors with rate $p_{\rm{unit}} = 0.1\%$ are introduced to data qubits.
    Throughout total physical time $T$, global decoding (blue, solid lines) is performed regularly for both original (blue circle) and hybrid protocol (orange diamond).
    %
    % Original and hybrid protocol performs it $N_{\rm{MWPM}}^{\rm{(Original)}}$ and $N_{\rm{MWPM}}^{\rm{(Hybrid)}}$ times, respectively.
    %
    However, hybrid decoding protocol regularly applies LEC circuit (orange, dotted lines) between global decoding applications.
    After $T$ passes, we evaluate the final logical error rate $P_{\rm{L}}$ (see Appendix \ref{appendix: error bar} for the error bar).
    %
    % (b) Method of finding a minimum number of global decoding for each protocol to achieve $P_{\rm{L}}}\ < 15$\% for fixed $T = 2000$ with 5000} samples.
    %
    % Our computation for the error bar of $P_{\rm{L}}$ is discussed in Appendix \ref{appendix: error bar}.
    %
    (b) Minimum number of global decoding vs. $p_{\rm{unit}} \times T$ in each protocol for given multi-qubit gate fidelities and $P_{\rm{L}}$ (see Appendix \ref{appendix: min-N-global} for details).
    } 
    \label{fig:4A_ReducingGlobal}
\end{figure}

% \bigskip\ptitle{Original vs. Hybrid protocol setting}

In practice, we aim for the memory to exhibit a logical error rate $P_{\rm{L}}$ below some target value after a given physical time $T$ where each unit time step introduces some error with rate $p_{\rm{unit}}$ to data qubits.
We compare original and hybrid decoding protocols as shown in Fig. \ref{fig:4A_ReducingGlobal}(a).
The original decoding protocol performs $N_{\rm{Global}}^{(\rm{Original})}$ global decoding with equally spaced time intervals throughout the total time $T$.
The hybrid decoding protocol performs $N_{\rm{Global}}^{(\rm{Hybrid})}$ global decoding with RL-optimized LEC circuits interleaved between global decoding.
%
% Note that each global decoding requires $L$ repetitions of mid-circuit readout of ancillas for faulty syndrome extraction before the 3D MWPM algorithm.
%
Our goal is to find the regime of parameters where
\begin{equation}\label{eq: N_MWPM_comparison}
    N_{\rm{Global}}^{\rm{(Original)}} > N_{\rm{Global}}^{\rm{(Hybrid)}}.
\end{equation}

% \bigskip\ptitle{Result comparison \& explanation}

When two-qubit gate fidelity $F_{\rm{2Q}} = 99.96\%$, three-qubit gate fidelity $F_{\rm{3Q}} = 99.88\%$, and target $P_{\rm{L}} < 15\%$, two protocols contrast as shown in Fig. \ref{fig:4A_ReducingGlobal}(b).
For $L = 8$ lattice, the hybrid decoding can reduce the number of global decoding for the constant amount regardless of $T$.
In other words, the \emph{gain} of using the hybrid decoding protocol gets decreased as we want to preserve the memory for a longer time.
However, for $L = 16$ lattice, we observe that the hybrid decoding can reduce the number of global decoding by about half regardless of how long we want to preserve the memory.
Note that each global decoding of 2D toric code requires $L$ repetitions of mid-circuit readout of ancillas for faulty syndrome extraction before the 3D minimum-weight perfect matching algorithm~\cite{Dennis_2002, wang2009threshold, higgott2023sparseblossomcorrectingmillion}.
Thus, we can conclude that using an LEC circuit reduces the number of mid-circuit readouts by $L\times(N_{\rm{Global}}^{\rm{(Original)}} - N_{\rm{Global}}^{\rm{(Hybrid)}})$ throughout the entire physical time, i.e. we can replace some portion of the mid-circuit readouts into possibly more cost-efficient operations.
Our investigation confirms that there exists a regime of high multi-qubit gate fidelity where LEC can be practically useful for a memory lifetime enhancement.

\subsection{Dissipatively preparing topological phases} \label{sec: VC}

% \ptitle{Standard preparation of states}

The same methods, developed here for suppressing errors in QEC codes, can be used for the systematic preparation of certain topological phases.
The logical subspace of topological error-correcting codes, such as the toric code, is a ground state of the corresponding commuting-projector Hamiltonian~\cite{Kitaev.2003}.
Such states can be prepared in constant time by measuring the stabilizers and then applying an appropriate correction to return to the ground state. 

% \bigskip\ptitle{Dissipative Preparation}

Alternatively, these ground states can be prepared by dissipation, with the circuit depth scaling as the linear size of the system~\cite{kalinowski_non-abelian_2023}.
The procedure described in this work can be interpreted as conditional transport of excitations, resulting in their eventual annihilation and a clean final state.
This stabilization (correction) procedure can be used as state preparation when applied to an appropriate initial state.

% \bigskip\ptitle{2D Toric Code Example}

Consider an example of a 2D toric code, starting in the symmetric product state.
This state is stabilized by the X checks but the Z checks have zero expectation value.
This can be interpreted as a condensate of the $m$-anyons, which need to be annihilated to prepare the target state.
Performing the conditional gates from our protocol amounts to inducing directional transport of these excitations, which eventually leads to their removal and lowering of state energy.
Our framework can be used to systematically construct and optimize such dissipative cooling protocols.

% \bigskip\ptitle{Further direction}

Extensions of these methods to more general error-correcting codes, whose parent Hamiltonians can exhibit exceedingly complicated low energy manifolds~\cite{anshu_nlts_2023}, is an interesting direction for future work.

\section{Discussion and Outlook} \label{sec: Discussion}

% \ptitle{Short summary of key results}

In this work, we develop an RL framework to optimize measurement-free error-correcting circuits, composed of a finite number of faulty multi-qubit gates.
We specifically focus on a finite-sized 2D toric code, 2D Ising model, and 4D toric code.
By introducing a notion of \emph{effective code distance} $D_{\rm{eff}}$, we find that when gate fidelities are high enough, the RL-optimized LEC circuits are better than conventional LEC circuits in extending the lifetime of a memory encoded in these codes.
Then, we additionally investigate two potential applications of this LEC scheme.
First, we show that the portion of global decoding, which consists of mid-circuit readout and classical decoding, can be replaced with LEC circuits with high-fidelity multi-qubit gates for preserving 2D toric code memory.
Also, we discuss that quantum states with topological phases can be prepared dissipatively.

Given an error-correcting code and gate error rate, our RL framework learns to optimize the decoding strategy as a sequence composed of operations from a restricted input gate set.
Since our task is to optimize such sequences for finite-sized systems, we find RL to be a more suitable tool compared to an analytical method discussed in other literature targeting systems in a thermodynamical limit~\cite{Brown_2016_QM, hong2024quantummemorynonzerotemperature}.

% \bigskip\ptitle{Outlook 1: Versatility}

Our RL framework is adaptive to the needs of various changes.
For example, keeping the same gate operations as an input action space, we can simply substitute a biased, realistic error model into our current noise channel.
Since coherent errors turn into Pauli errors when we trace out the ancillas at the ancilla reset step~\cite{Bravyi_2018, Iverson_2020, zhao2021analyticstudyindependentcoherent, Marton2023coherenterrors}, our RL framework can be used for correcting coherent physical errors in the code as well.
Moreover, we can extend our action space by introducing additional error removal operations that can correct higher-weight errors.
It is also possible to apply our RL framework to different codes not discussed in this work as well.
For example, it would be interesting to consider higher-dimensional color codes, which support transversal non-Clifford logical operations~\cite{Kubica_2015, Haah_2021, Honciuc_Menendez_2024}, have interesting single-shot error correction properties~\cite{Bomb_n_2015, Brown_2016}, and can also be realized in state-of-the-art quantum platforms~\cite{Bluvstein_2023, dasilva2024demonstration}.

% \bigskip\ptitle{Outlook 2: More complicated RL structure}

Although our RL framework benefits from a simple and small neural network in terms of the training cost, we can choose a more complicated optimizer --- including transformer~\cite{vaswani2023attention} that has already been used in designing an efficient \emph{global} decoder for 2D surface code~\cite{bausch2023learning, wang2023transformerqec}.
% However, Acknowledge no unique advantage!
%
The advantage of using these more advanced schemes is that we may not need to provide a fixed gate set --- depending on our knowledge of error patterns on a given error-correcting code --- as input.
We can allow RL to find the proper gate input set \emph{from scratch} by only providing the type of available gates as input ingredients.
This future direction can also be useful to obtain a \emph{general} LEC circuit optimizer with the stabilizer structure of the code as an input to the RL agent such that we do not have to re-train the agent for each code.

\begin{acknowledgements}
    We thank Aur\'elien Dersy, Yanke Song, Lucas Janson, Timothy H. Hsieh, Tsung-Cheng Lu, Shengqi Sang, Hong-Ye Hu, Di Luo, Andrew K. Saydjari, Kevin B. Xu, Minhak Song, Sangheon Lee, Jin Ming Koh, and Juan Pablo Bonilla Ataides for insightful discussions.
    We also thank Aur\'elien Dersy for a critical reading of the manuscript.
    We utilize the Stable-Baselines3 package for implementing the RL PPO algorithm~\cite{stable-baselines3}, PyMatching package for implementing the MWPM algorithm for 2D toric code~\cite{higgott2021pymatching, higgott2023sparseblossomcorrectingmillion}, and the symfit package for fitting multiple datasets~\cite{Roelfs2017}.
    Our visualization of 2D toric code and 2D Ising model is motivated by the codebase of previous works~\cite{Andreasson_2019, Sweke_2020, Fitzek_2020}.
    M.P. acknowledges funding from Herchel Smith-Harvard Undergraduate Science Research Program (Summer 2022), Harvard Quantum Initiative Undergraduate Research Fellowship (Summer 2022), and Harvard College Research Program (Fall 2022).
    N.M. acknowledges support from the Department of Energy Computational Science Graduate Fellowship under award number DE-SC0021110.
    We acknowledge the financial support by the Center for Ultracold Atoms (an NSF Physics Frontiers Center), Wellcome LEAP, LBNL/DOE QSA (grant number DE-AC02-05CH11231), NSF (grant number PHY-2012023), DARPA ONISQ (grant number W911NF2010021), DARPA IMPAQT (grant number HR0011-23-3-0030), DARPA MeasQuIT (grant number HR0011-24-9-0359), and IARPA ELQ (grant number W911NF2320219).
\end{acknowledgements}

% \clearpage
% \pagebreak

\appendix

\section{Experimental implementation of LEC} \label{appendix: exp}

One of the possible limitations of the LEC circuit is that we regularly reset the ancillas to initial states $\ket{0}$ before we perform a new syndrome extraction operation.
In certain state-of-the-art quantum devices, the ancillas with errors can be reset to the initialized ones --- regardless of their quantum states --- through ancilla repumping techniques~\cite{PhysRevX.13.041035, PhysRevX.13.041051}.
Thus, employing these techniques, it is possible for us to perform multiple LEC cycles with a finite number of qubits.
Note that the advantage of using the LEC scheme over the standard error correction scheme with ancilla readout is available only if the local coherent operations are more efficient compared to the mid-circuit readout.
Thus, an adaptation of this LEC scheme in the actual quantum device should require a careful analysis of resources available within certain quantum platforms.

Moreover, a layout of which gates are applied to which qubits is specific to each syndrome extraction and error removal operation.
Since qubits must be adjacent to each other to perform multi-qubit gates, we change the spatial configuration of data qubits and ancillas before applying each gate for a syndrome extraction and error removal operation.
Note that the qubits can be moved coherently and independently in several state-of-the-art quantum processors, including neutral-atom array platforms~\cite{Bernien_2017, PhysRevApplied.19.034048, Xu2024}.
Thus, each operation consists of a qubit reconfiguration operation followed by corresponding multi-qubit gates.
For simplicity, we ignore the errors from the process of this dynamical rearrangement of qubits and suppose that the qubit reconfiguration operation is perfect.

\begin{table*}[t]
\renewcommand{\arraystretch}{1.5}
\begin{tabular}{ |c||c|c|c||c|c|c| } 
 \hline
  & Control 1 of CCX & Control 2 of CCX & Target of CCX & Control 1 of CCZ & Control 2 of CCZ & Target of CCZ \\ 
 \hline\hline
 Type 1 & $(o,e,e,e)$ & $(e,o,e,e)$ & $(o,o,e,e)$ & $(e,o,o,o)$ & $(o,e,o,o)$ & $(e,e,o,o)$ \\
 \hline
 Type 2 & $(o,e,e,e)$ & $(e,e,o,e)$ & $(o,e,o,e)$ & $(e,o,o,o)$ & $(o,o,e,o)$ & $(e,o,e,o)$ \\
 \hline
 Type 3 & $(o,e,e,e)$ & $(e,e,e,o)$ & $(o,e,e,o)$ & $(e,o,o,o)$ & $(o,o,o,e)$ & $(e,o,o,e)$ \\
 \hline
 Type 4 & $(e,o,e,e)$ & $(e,e,o,e)$ & $(e,o,o,e)$ & $(o,e,o,o)$ & $(o,o,e,o)$ & $(o,e,e,o)$ \\
 \hline
 Type 5 & $(e,o,e,e)$ & $(e,e,e,o)$ & $(e,o,e,o)$ & $(o,e,o,o)$ & $(o,o,o,e)$ & $(o,e,o,e)$ \\
 \hline
 Type 6 & $(e,e,o,e)$ & $(e,e,e,o)$ & $(e,e,o,o)$ & $(o,o,e,o)$ & $(o,o,o,e)$ & $(o,o,e,e)$ \\
 \hline
\end{tabular}
\caption{Different configurations of parallelized Toom Sweep operations. Note that each operation consists of $L^2$ number of CCX gates and $L^2$ number of CCZ gates. Also, each configuration has four different ``directions.'' For example, a CCX gate targeted to data qubit on $(1,1,2,2)$ can be controlled by $\hat{Z}$ stabilizer ancilla on either $(1\pm 1, 1, 2, 2)$ and either $(1, 1\pm 1, 2, 2)$.}
\label{tab:4dToric-Toom}
\end{table*}
\renewcommand{\arraystretch}{1}

\begin{table}[t]
\renewcommand{\arraystretch}{1.5}
\begin{tabular}{ |c||c||c| } 
 \hline
 & CCX (Ctrls $-$ Targ) & CCZ (Ctrls $-$ Targ) \\ 
 \hline\hline
 & $(o,e,e,e)-(o,o,e,e)$ & $(o,e,o,o)-(e,e,o,o)$ \\ 
 Type 1 & $(e,o,e,e)-(e,o,o,e)$ & $(o,o,o,e)-(o,e,o,e)$ \\
 & $(e,e,e,o)-(o,e,e,o)$ & $(e,o,o,o)-(e,o,e,o)$\\
 \hline
 & $(e,o,e,e)-(o,o,e,e)$ & $(e,o,o,o)-(e,e,o,o)$ \\ 
 Type 2 & $(e,e,o,e)-(e,o,o,e)$ & $(o,e,o,o)-(o,e,o,e)$ \\
 & $(o,e,e,e)-(o,e,e,o)$ & $(o,o,e,o)-(e,o,e,o)$\\
 \hline
 & $(e,e,o,e)-(e,e,o,o)$ & $(o,o,e,o)-(o,o,e,e)$ \\ 
 Type 3 & $(e,e,e,o)-(e,o,e,e)$ & $(o,e,o,o)-(o,e,e,o)$ \\
 & $(o,e,e,e)-(o,e,o,e)$ & $(o,o,o,e)-(e,o,o,e)$\\
 \hline
 & $(e,e,e,o)-(e,e,o,o)$ & $(o,o,o,e)-(o,o,e,e)$ \\ 
 Type 4 & $(e,o,e,e)-(e,o,e,o)$ & $(o,o,e,o)-(o,e,e,o)$ \\
 & $(e,e,o,e)-(o,e,o,e)$ & $(e,o,o,o)-(e,o,o,e)$\\
 \hline
\end{tabular}
\caption{Different configurations of parallelized $d=1$ and $d=(1,1)$ operations. Note that each operation consists of $3L^2/2$ number of CCX gates and $3L^2/2$ number of CCZ gates, and control ancillas are on the same parity classifications of coordinates.}
\label{tab:4dToric-dN}
\end{table}
\renewcommand{\arraystretch}{1}

\section{Circuit-level simulation of QEC} \label{appendix: QECC-implementation}

In this section, we discuss how we implement the circuit-level simulation of QEC in each code: 2D toric code, 2D (classical) Ising model, and 4D toric code.
Note that for 2D toric code and 2D Ising model, our visualization of the lattice, data qubits, and syndromes as in Fig. \ref{fig:1A-1_StabilizerDefinition} and \ref{fig:1B_Uncorrectable} are modified based on the codebase of previous works~\cite{Andreasson_2019, Fitzek_2020}.

We perform stabilizer-level Monte Carlo simulation based on NumPy for each code, and our error model includes unbiased bit-flip and dephasing errors:
\begin{itemize}
    \item We represent data qubits as three-dimensional arrays: first dimension for a type of a Pauli error on the qubit, second dimension for an index of a copy of a code, and third dimension for an index of a qubit within a code.

    \item To perform a parallelized CNOT, CCX, or CCZ gate operation, we define lists of qubit indices that match with the gate connectivity.
    The $i$-th index for each list corresponds to control or target qubits for the $i$-th multi-qubit gate to be applied.
    % For example, for parallelized CCX gates, we define two lists for ancilla indices and a list for data qubit indices.
    %
    % Then, the $i$-th index for each list corresponds to control and target qubits for the $i$-th CCX gate to be applied.
    %
    Although we implement the Pauli error propagation for each CNOT gate, we model each ancilla classically, i.e. either $\ket{0}$ or $\ket{1}$, to avoid dealing with error propagation through non-Clifford CCZ and CCX gates.
    Physically, this condition can be approximately achieved by resetting ancillas regularly as illustrated in Fig. \ref{fig:1_Overview}.

    % \item When ambient or gate errors with error rate $p_{\rm{err}}$ are introduced to the qubits, we generate other arrays with the same shapes where each entry is 0 with probability $1-p_{\rm{err}}$ and 1 with probability $p_{\rm{err}}$ and then perform XOR operation between the original arrays and such ``error arrays.''
    % %
    % Note that for each copy of code, different patterns of errors are introduced to data qubits and ancillas.
    % %
    % This enables us to benchmark a decoding success rate of the given LEC circuit with given error rates.
\end{itemize}

We define each parallelized $d=N$ gate or Toom Sweep gate as a tessellation of the unit cell as illustrated in Fig. \ref{fig:1C_ExtendedOperations}.
For 2D toric code and 2D Ising model, the visualization is straightforward as shown in Fig. \ref{fig:1A-2_CircuitOperations}.
For 2D toric code, we include four $d=1$, eight $d=2$, and twelve $d=3$ operations with different unit cells as the candidates of error removal operations.
For the 2D Ising model, we include four Toom Sweep operations (with $90^\circ$ rotation), four $d=1$ operations, and four $d=2$ operations as the candidates of error removal operations.

However, for 4D toric code, we effectively represent the data qubits and ancillas using odd and even coordinates~\cite{PhysRevResearch.3.013118}.
Let's consider a coordinate $(x_0,x_1,x_2,x_3)$ on 4D lattice, where each $x_i$ has an integer value between 0 and $2L-1$.
%
% Then, we consider data qubits --- on ``faces'' of the 4D lattice  --- are located on coordinates where two of $x_i$s are even and two of them are odd.
%
% Similarly, we consider ancillas that store eigenvalues of edge-type (or cube-type) stabilizers with six $\hat{Z}$ (or $\hat{X}$) operators to be located on coordinates where three of $x_i$s are even (or odd) and one of them is odd (or even).
%
We denote odd coordinates as $o$ and even coordinates as $e$ to classify qubits:
\begin{itemize}
    \item Data qubits are on $(o,o,e,e)$, $(o,e,o,e)$, $(o,e,e,o)$, $(e,o,o,e)$, $(e,o,e,o)$, and $(e,e,o,o)$ --- $6L^2$ in total.
    \item $\hat{Z}$ stabilizer ancillas are on $(o,e,e,e)$, $(e,o,e,e)$, $(e,e,o,e)$, and $(e,e,e,o)$ --- $4L^2$ in total.
    \item $\hat{X}$ stabilizer ancillas are on $(e,o,o,o)$, $(o,e,o,o)$, $(o,o,e,o)$, and $(o,o,o,e)$ --- $4L^2$ in total.
\end{itemize}
% This representation provides an intuitive understanding of which data qubits support each stabilizer.
%
% For example, an edge-type stabilizer on $(1,2,2,2)$ is made of $\hat{Z}$ operators on data qubits $(1,1,2,2)$, $(1,2,1,2)$, $(1,2,2,1)$, $(1,3,2,2)$, $(1,2,3,2)$, and $(1,2,2,3)$.
%
% Also, such structure of connectivity provides us an understanding of logical $\hat{Z}$ and $\hat{X}$ operators.
From this representation, we can easily identify six independent logical $\hat{Z}$ and $\hat{X}$ operators.
For example, a topologically non-trivial ``sheet'' of $\hat{Z}$ operators on all data qubits with coordinates $(1,1,2k_1,2k_2)$ with $0 \le k_1, k_2 \le L-1$ correspond to logical $\hat{Z}$ operator, which indeed commutes with all stabilizers.
%
% From this analysis, six independent logical $\hat{Z}$ and $\hat{X}$ operators can be easily identified.

Now, we introduce error removal operations based on CCX and CCZ gates in parallel for 4D toric code.
As shown in Table \ref{tab:4dToric-Toom}, there are six different configurations of parallelized Toom Sweep operations, where each consists of $L^2$ CCX gates and $L^2$ CCZ gates.
Since each configuration has four choices of ``direction'' just as in the 2D Ising model illustrated in Fig. \ref{fig:1C_ExtendedOperations}(b), we have $6 \times 4 = 24$ parallelized Toom Sweep operations.
On the other hand, CCX/CCZ gates for both $d=1$ and $d=(1,1)$ operations are controlled by the same parity classification of ancillas.
As shown in Table \ref{tab:4dToric-dN}, there are four different configurations of parallelized $d=1$ and $d=(1,1)$ operations, where each consists of $3L^2/2$ CCX gates and $3L^2/2$ CCZ gates --- this is why we choose $L$ to be even.
For $d=1$ operation, each configuration has two choices for a group of data qubits.
%
% For example, $d=1$ CCX targeted to $(1,1,2,2)$ and controlled by $(1,0,2,2)$ and $(1,2,2,2)$ cannot be done in parallel with $d=1$ CCX targeted to $(1,3,2,2)$ and controlled by $(1,2,2,2)$ and $(1,4,2,2)$.
%
Thus, we have $4 \times 2 = 8$ parallelized $d=1$ operations.
For $d=(1,1)$ operation, there are four additional degrees of freedom on top of two choices for a group of data qubits.
%
% For example, $d=1$ CCX targeted to $(1,1,2,2)$ can be controlled by $(1,0,2,2)$ and either $(1,2,0,2)$, $(1,2,4,2)$, $(1,2,2,0)$, or $(1,2,2,4)$.
%
Thus, we have $4 \times 2 \times 4 = 32$ parallelized $d=(1,1)$ operations.
In summary, we include twenty-four Toom Sweep operations, eight $d=1$ operations, and thirty-two $d=(1,1)$ operations as the candidates of error removal operations.
For simplifying the RL training and reducing the number of syndrome extraction operations, we ``grouped'' parallelized $d=1$ and $d=(1,1)$ operations each into single actions.

From such design of error removal operations, we optimize the order of applying parallelized CNOT gates for syndrome extraction operation.
For 2D toric code, there are four ``steps'' of parallelized CNOT gates in sequence to perform a syndrome extraction as illustrated in Fig. \ref{fig:1A-2_CircuitOperations}.
For the 2D Ising model and 4D toric code, we perform a syndrome excitation before each error removal operation.
We only update syndromes for the ancillas that would be used in the following error removal operation.
%
% For example, for Type 1 Toom operation in Table \ref{tab:4dToric-Toom}, we only prepare stabilizer ancillas on $(o,e,e,e)$, $(e,o,e,e)$, $(e,o,o,o)$, $(o,e,o,o)$, i.e. $4L^2$ ancillas instead of $8L^2$.
%
% We decompose such syndrome extraction for $4L^2$ ancillas into six ``steps'' of parallelized CNOT gates in sequence.
%
After finding the optimal $N$ number of steps of parallelized CNOT gates in sequence, we test $N!$ possible permutations of this sequence and choose the one with the smallest average number of errors introduced during the entire syndrome extraction operation.
Note that it is also possible to optimize such sequence more systematically as discussed in previous works~\cite{PhysRevA.101.012342, ryananderson2022implementingfaulttolerantentanglinggates}.

\section{Details on RL training} \label{appendix: RL}

\subsection{Details on agent} \label{appendix: RL-agent}

We summarize the RL agent's construction of an LEC circuit as in Algorithm \ref{alg: agent}, which defines the ``BuildLEC'' function in Algorithm \ref{alg: RL}.

\begin{algorithm}
\caption{``BuildLEC'' (RL agent)}\label{alg: agent}
\SetKwFunction{FMain}{BuildLEC}
    \KwData{Neural network with parameters $f^{\vec{\theta}}$; circuit depth $H$; action space $\mathcal{A}$}
    \SetKwProg{Fn}{Function}{:}{}
    \Fn{\FMain{$f^{\vec{\theta}}$, $H$, $\mathcal{A}$}}{
        Circuit $\gets$ EmptyList\;
        \For{$ h < H $}
        {
        NewAction := $f^{\vec{\theta}}$(Circuit)\;
        Circuit.insert(NewAction)
        }
        \textbf{return} Circuit;
}
\end{algorithm}

The size of the action space $|\mathcal{A}|$ depends on the code: $|\mathcal{A}| = 24$ for 2D toric code, $|\mathcal{A}| = 12$ for 2D Ising model, and $|\mathcal{A}| = 26$ for 4D toric code (see Appendix \ref{appendix: QECC-implementation}).
Also, in practice, we use two fully connected hidden layers with 128 nodes each for both policy and value networks.
We provide a justification of not using bigger networks in Appendix \ref{appendix: RL-hyperparameters}.
% Note that both types are based on the scheme that given a circuit depth $H$, the RL agent builds a sequence of $H$ actions as illustrated in Fig. \ref{fig:2A_SummaryRL}(a).

% A neural network $f^{\vec{\theta}}$ in the agent takes \emph{all previous} actions as an input and outputs the next action.
%
To represent the previous actions as an observation for the RL agent, we one-hot encoded the LEC action sequence into $|\mathcal{A}|$-by-$H$ matrix with entries $0$ or $1$.
%
% For example, the LEC action sequence $\{A_1, \ldots, A_h\}$ can be written by the $|\mathcal{A}|$-by-$H$ matrix with $(A_h,h)$ entries are $1$ and all other entires are $0$.
%
For 2D toric code, each action $A_i$ is either syndrome extraction or one of error removal operations in Fig. \ref{fig:1C_ExtendedOperations}(a), and we fix that the first action $A_1$ is always a syndrome extraction operation.
For the 2D Ising model and 4D toric code, each action $A_i$ is one of the error removal operations in Fig. \ref{fig:1C_ExtendedOperations}(b) followed by syndrome extraction operation on corresponding ancillas.
We do not fix the first action for the 2D Ising model and 4D toric code, because a syndrome extraction operation is performed for any action before an error removal operation.
%
% We optimize and fix the order of CNOT gates that consist of each syndrome extraction operation as discussed in Appendix \ref{appendix: QECC-implementation}.

As discussed in Section \ref{sec: IIIB}, we construct two different types of RL agents: (1) fixed-depth agent for 2D Ising model and 4D toric code and (2) variable-depth agent for 2D toric code.
For a fixed-depth agent, we consider $H$ to be the fixed LEC circuit depth.
However, for a variable-depth agent, we modify the above scheme such that it optimizes the circuit depth $H \le H_{\max}$ where $H_{\max}$ is a fixed input parameter.
To achieve this, we additionally include a \emph{null} action $\emptyset$ where $A_h = \emptyset$ implies that the $h$-th LEC action is ``skipped'' within a sequence of $H_{\max}$ actions.
%
% We consider a new action space $\mathcal{A}' = \mathcal{A} \cup \{\emptyset\}$ and perform the RL training with such $\mathcal{A}'$ and fixed circuit depth $H_{\max}$.
%
Then, removing all $\emptyset$s from the RL-optimized LEC circuit with $H_{\max}$ results in the RL-optimized LEC circuit with $H_{\rm{opt}}$ as desired.
%
% By comparing the performance between the LEC circuit with $H_{\rm{opt}}$ and fixed $H$ in Fig. \ref{fig:2B_OptimizedDepthLayout}(a), we confirm that $H_{\rm{opt}}$ is indeed an optimal length of the LEC circuit for fixed input training $p_{\rm{gate}}$.

Two limitations of this modified RL framework are that (1) $H_{\rm{opt}} \le H_{\max}$ and (2) $H_{\rm{opt}}$ found by RL training is suboptimal if true $H_{\rm{opt}}$ is much smaller than $H_{\max}$.
We set initial $H_{\max}$ to be 40 for 2D toric code as we confirmed from training with the fixed-depth agent that $H_{\rm{opt}}$ is smaller than 40.
Then, after the termination condition is satisfied and $H_{\rm{opt}}$ is obtained, we set the next $H_{\max}$ to be the $H_{\rm{opt}}$ of the previous training and run the next training.
This is why we emphasize in Section \ref{sec: IIIB} that the training with a variable-depth agent is more ``expensive'' than the training with a fixed-depth agent.

We have tried an alternative approach to optimize the LEC circuit with a variable depth by adding a \emph{stop} action that terminates the LEC circuit immediately --- instead of the $\emptyset$ action.
%
% a \emph{stop} action --- instead of the null action $\emptyset$.
%
% Instead of ``doing nothing,'' this action terminates the LEC circuit immediately.
%
This approach was not successful, because the RL training resulted in building the NN LEC circuit with circuit depth 5~\cite{Smith_2023}, even for training $p_{\rm{gate}} = 1 \times 10^{-4}$ where a true $H_{\rm{opt}} > 30$.
Since the NN LEC circuit is good enough to possibly form a ``deep'' local minimum in the reward landscape, we believe that a \emph{stop} action is incentivized enough to be called early.
Although it could be possible to get around this issue by modifying the reward structure, we find that the $\emptyset$-action option is natural to achieve the variable-depth LEC circuit optimization.

\subsection{Details on environment}\label{appendix: RL-env}

We summarize the RL environment's computation of reward as in Algorithm \ref{alg: env}, which defines the ``ComputeReward'' function in Algorithm \ref{alg: RL}.

We first run a simulation that applies multiple rounds of LEC.
In each round, errors are introduced to data qubits, and then the input LEC circuit (fixed across all rounds) is applied to data qubits and newly initialized ancillas.
Note that we are interested in extending the lifetime of memory encoded in the error-correcting code, and local decoding is performed multiple times through the lifetime of memory.
Thus, we aim to reduce the \emph{accumulated} error density over multiple rounds.
Note that RL-optimized circuits used in Fig. \ref{fig:2B_OptimizedDepthLayout}, \ref{fig:3C-1_FittingDeff}, and \ref{fig:3A_ErrorDistribution} are trained with the code and error parameters in Table \ref{tab: parameters} --- $p_{\rm{gate}}$ varies for each training.

\begin{table}[h]
\renewcommand{\arraystretch}{1.5}
\begin{tabular}{ |c||c|c|c|c| }
 \hline
 & $L$ & $p_{\rm{amb}}$ & $N$ & $R$ \\ 
 \hline\hline
 2D toric code & 8 & 0.02 & 100 & 5 \\
 \hline
 2D Ising model & 8 & 0.40 & 100 & 1 \\ 
 \hline
 4D toric code & 4 & 0.03 & 50 & 2 \\
 \hline
\end{tabular}
\caption{Code and error model parameters used for optimizing the LEC circuit with our RL framework.
$L$ denotes the system size, $p_{\rm{amb}}$ denotes the ambient error rate in each LEC cycle before applying the LEC circuit, $N$ denotes the number of copies of code, and $R$ denotes the number of repeated LEC cycles used to compute decoding success rate as illustrated in Fig. \ref{fig:2A_SummaryRL}.
}
\label{tab: parameters}
\end{table}

\begin{algorithm*}
\caption{``ComputeReward'' (RL env.)}\label{alg: env}
\KwData{Code type $T$; lattice size $L$; ambient error rate $p_{\rm{amb}}$; gate error rate $p_{\rm{gate}}$; sample size $N$}
\SetKwFunction{FMain}{ComputeReward}
    \SetKwProg{Fn}{Function}{:}{}
    \Fn{\FMain{circuit, $T$, $L$, $p_{\rm{amb}}$, $p_{\rm{gate}}$, $N$}}{
        result $\gets$ EmptyList\;
        \For{sample $\in \{1, \ldots, N\} $}
        {
        data $\gets$ initialized data qubits\;
        \For{round $\in \{ 1, \ldots, R \}$}{
            data := GenerateError(data, $p_{\rm{amb}}$, $T$)\;
            ancillas $\gets$ initialized ancillas\;
            \For{gate $\in$ circuit}{
                data, ancillas := ApplyGate(data, ancillas, gate)\;
                $\{$data, ancillas$\}$ := GenerateError($\{$data, ancillas$\}$, $p_{\rm{gate}}$, $T$)\;
            }
        }
        RewardEach = EvaluateRewardEach(data, $T$)\;
        result.insert(RewardEach);
        }
        \textbf{return} Average of result;
}

\SetKwFunction{FMain}{GenerateError}
    \SetKwProg{Fn}{Function}{:}{}
    \Fn{\FMain{qubits, $p_{\rm{err}}$}}{
        \For{qubit $\in$ qubits}{
            randX $\in$ Uniform([0,1])\;
            \If{randX $< p_{\rm{err}}$}{
                Apply $\hat{X}$ error on qubit\;
              }
            
            \If{$T$ is 2D or 4D toric code}{
                randZ $\in$ Uniform([0,1])\;
                \If{randZ $< p_{\rm{err}}$}{
                    Apply $\hat{Z}$ error on qubit\;
                }
              }
        }
        \textbf{return} qubits;
}

\SetKwFunction{FMain}{EvaluateRewardEach}
    \SetKwProg{Fn}{Function}{:}{}
    \Fn{\FMain{data, $T$}}{
        RewardEach $\gets$ 0\;
        \uIf{$T$ is 2D toric code}{
            Perform MWPM algorithm on the precise eigenvalues of all stabilizers\;
            \If{Resulting state is a trivial ground state of 2D toric code without logical error}{RewardEach := 1}
          }
          \uIf{$T$ is 2D Ising model}{
            RewardEach := 1 $-$ (Number of flipped spins among data qubits) / $L^2$
          }
          \If{$T$ is 4D toric code}{
            Perform $\ge 50$ repetitions of perfect Toom's rule decoding\;
            \If{Resulting state is a trivial ground state of 4D toric code without logical error}{RewardEach := 1}
          }
        \textbf{return} RewardEach;
}
\end{algorithm*}

%
% We consider two different error sources: 
% \begin{itemize}
    % \item \emph{Ambient} errors introduced to \emph{all} data qubits every LEC round before applying LEC circuit with error rate $p_{\rm{err}} = p_{\rm{amb}}$,
    % \item \emph{Gate} errors introduced to \emph{all} data qubits and ancillas after each parallelized multi-qubit gate operation with error rate $p_{\rm{err}} = p_{\rm{gate}}$.
% \end{itemize}
% We adopt an unbiased error model with bit-flip and dephasing errors.
%
% For 2D and 4D toric code, we apply $\hat{X}$ and $\hat{Z}$ error independently to each qubit with the same probability of $p_{\rm{err}}$ ($\hat{Y} \sim \hat{X}\hat{Z}$ error applied with a probability of $p_{\rm{err}}^2$).
%
% For the 2D Ising model, we apply $\hat{X}$ error independently to each qubit with the probability of $p_{\rm{err}}$.

%% Decoding success evaluation
After multiple LEC rounds, we evaluate the decoding success.
This \emph{final recovery} step is to determine whether the final state of data qubits after multiple LEC rounds is ``close enough'' to the trivial ground state of the error-correcting code.
This step varies among the choice of the code.
For 2D toric code, we perform global decoding by minimum-weight perfect matching (MWPM) classical algorithm with precise syndrome information using PyMatching package~\cite{higgott2021pymatching, higgott2023sparseblossomcorrectingmillion}.
For 2D Ising model, we perform a majority vote among data qubits.
For 4D toric code, we perform enough (50 rounds) repetition of perfect Toom's rule decoding.
After this step, 2D and 4D toric code decoding are evaluated to be successful if we get a trivial ground state, whereas 2D Ising model decoding is evaluated to be successful if we get less than 50\% of all spins bit-flipped.
We repeat this simulation for $N$ times with a fixed LEC circuit.
For each simulation, a stochastic error model results in different configurations of errors applied to qubits.
Thus, we can obtain the decoding success rate through this benchmark simulation.

As discussed in Section \ref{sec: IIB}, the uncorrectable error configurations after perfect Toom's rule decoding in 4D toric code can have a low error density.
In principle, these uncorrectable errors should be decoded by some non-local decoding protocol~\cite{Breuckmann_2018}.
However, we decide to leave these uncorrectable errors as decoding failures, because (1) the non-local decoding protocol takes significant running time and (2) the prior LEC reduces the uncorrectable error configurations with low error density.

For 2D and 4D toric code, the reward is computed as the number of successful decoding cases divided by the number of total cases $N$.
However, for the 2D Ising model, the reward is computed as an average non-error density, i.e. average number of non-flipped spins divided by the total number of spins $L^2$.
%
% This reward computation method is summarized in Algorithm \ref{alg: env}.

% Also, we choose specific parameters of the RL environment in each code as shown in Table \ref{tab: parameters}.
%
% Lastly, we choose the parameters in Table \ref{tab: parameters} based on comparing the final rewards from training with various choices of this parameter set.
%
In particular, we choose $R$ in  Table \ref{tab: parameters} to be as small as possible for each code to minimize the time for reward computation.
However, for 2D toric code, we observe that $R \le 4$ causes the occurrence of repeating syndrome extraction, which introduces more gate errors into the system without benefiting decoding.
This is because, for such a small $R$, the RL agent is not penalized enough by a bad reward to avoid repeating syndrome extraction and to choose the ``do nothing'' ($\emptyset$) action instead.
Thus, we choose $R = 5$ for the 2D toric code case.

\subsection{Conditions on training termination} \label{appendix: stop-training}

Each epoch of our RL training consists of 500 episodes with a mini-batch size of 50.
In other words, we collect 500 samples of LEC circuits and their benchmarked decoding success rates, divide them into smaller mini-batch subsets with 50 samples each, and perform 10 updates with these subsets on the policy and value network.
Before saving our updated neural network model, we run 40 epochs for the 2D toric code, 80 epochs for the 2D Ising model, and 40 epochs for the 4D toric code.
We terminate each RL training when a neural network model produces the same LEC circuit as output as the previously saved one throughout these epochs.
For 2D toric code with $H_{\max} = 40$ and $p_{\rm{gate}} = 1 \times 10^{-4}$, each RL training requires 200-400 epochs.
For self-correcting memories with $H = 60$ and any $p_{\rm{gate}}$, each RL training requires 800-1000 epochs for the 2D Ising model and 400-600 epochs for the 4D toric code.
Note that these are obtained with a fully connected neural network with two hidden layers of size 128.
Also, we use a variable depth agent for 2D toric code and a fixed depth agent for 2D Ising model and 4D toric code (see Appendix \ref{appendix: RL-agent}).

\subsection{Details on RL hyperparameters} \label{appendix: RL-hyperparameters}

We utilize the initial given values of hyperparameters for the PPO algorithm in Stable-Baselines3 library~\cite{stable-baselines3} with the following exceptions:
\begin{itemize}
    \item Since we compute the reward only at the end of each episode, we set a discount factor $\gamma$ to be 1.

    \item n\_steps and batch\_size parameters are set to be $500 \times H$ and $50 \times H$, where $H$ is $H_{\max}$ for 2D toric code and a fixed circuit depth for 2D Ising model and 4D toric code.

    \item We use a fully connected neural network with two hidden layers of size 128 for both policy and value networks.
    The reasoning for this choice is based on the following paragraph.
\end{itemize}

As shown in Table \ref{tab:varyNN}, we compare the final reward from RL training with a fully connected neural network with two hidden layers of size $d_{\rm{NN}}$ with four independent runs.
We observe that the final reward gets improved until $d_{\rm{NN}} = 128$.
However, for $d_{\rm{NN}} > 128$, we observe that a similar reward was obtained with shorter training epochs before termination.
Considering the size of saved neural networks after training, we choose $d_{\rm{NN}} = 128$ for our RL training.
Also, this is why we did not have to implement a more complicated neural network structure --- such as a transformer --- because a simple fully connected neural network with two hidden layers of size 128 was enough for our optimization task.

We attempt the optimization of other hyperparameters --- such as learning\_rate or n\_epochs --- via Optuna framework~\cite{akiba2019optunanextgenerationhyperparameteroptimization}.
However, we find that our RL training has an expensive reward computation and thus not ideal for such hyperparameter optimization.
Further, we observe that even with given hyperparameters by Stable Baselines 3, we obtain a satisfactory optimization of the LEC circuits as discussed in Section \ref{sec: III} and \ref{sec: IV}.
Therefore, we did not further investigate such optimization of other RL hyperparameters besides the network parameters.

\subsection{Stability of RL Training} \label{appendix: RL-stability}

\begin{figure}[b]
    \centering
    \includegraphics[width=\linewidth]{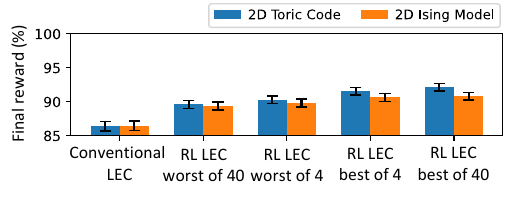}
    \caption{Final reward comparison of conventional vs. RL LEC chosen from worst or best runs out of multiple training executions.
    Note that conventional LEC is a nearest-neighbor LEC circuit for 2D toric code and Toom's rule-based LEC circuit for 2D Ising model and reward is computed out of 10000 independent samples.
    RL training for 2D toric code and 2D Ising model are done with $p_{\rm{gate}} = 1\times 10^{-4}$ and $1\times 10^{-3}$, respectively, and with other parameters in Table \ref{tab: parameters}.}
    \label{fig:RL-stability}
\end{figure}

As discussed in Section \ref{sec: IIIA}, we choose one LEC circuit out of 4 RL executions that maximizes the reward.
We record rewards by RL LECs from 4 runs as well as the reward by the conventional LEC with the same parameters as in Table \ref{tab: RL-variations}.
Also, we perform 40 runs of RL training for 2D toric code and 2D Ising model with $p_{\rm{gate}} = 1 \times 10^{-4}$ and $1 \times 10^{-3}$, respectively.
As shown in Fig. \ref{fig:RL-stability}, we record the maximum and minimum rewards by RL LECs out of the first 4 and all 40 runs for each code as well as the reward by the conventional LEC with the same parameters.
The final reward of any RL LEC is higher than that of conventional LEC, which we benchmark the decoding performance against.
Also, we find that the best final reward out of 4 and 40 runs is marginally different, which justifies the number of RL training executions.
These results support the stability of RL training, i.e., RL training results do not vary significantly in performance with respect to our demand.

\begin{table*}[t]
\renewcommand{\arraystretch}{1.5}
\begin{tabular}{ |c||c|c|c|c||c||c|c|c|c| } 
 \hline
 \multicolumn{5}{|c||}{2D Toric Code} & \multicolumn{5}{c|}{2D Ising Model} \\
 \multicolumn{5}{|c||}{$(L, H_{\max} N, R, p_{\rm{amb}}, p_{\rm{gate}}) = (8, 40, 100, 7, 0.02, 1\times 10^{-4})$} & \multicolumn{5}{c|}{$(L, H_{\max} N, p_{\rm{amb}}, p_{\rm{gate}}) = (8, 60, 100, 0.40, 1\times 10^{-3})$} \\
 \hline
 $d_{\rm{NN}}$ & Run 1 & Run 2 & Run 3 & Run 4 & $d_{\rm{NN}}$ & Run 1 & Run 2 & Run 3 & Run 4 \\ 
 \hline\hline
 32 & 0.1440 (440) & 0.1668 (480) & 0.1463 (520) & 0.1467 (520) & 32 & 0.1022 (1520) & 0.1011 (1600) & 0.0973 (1360) & 0.0981 (1520)\\
 \hline
 64 & 0.1668 (280) & 0.1468 (360) & 0.1489 (400) & 0.1634 (320) & 64 & 0.1015 (1120) & 0.0975 (1360) & 0.1021 (1120) & 0.0986 (1120)\\
 \hline
 128 & 0.1476 (280) & 0.1505 (240) & 0.1498 (240) & 0.1263 (280) & 128 & 0.0981 (1120) & 0.1080 (1040) & 0.0973 (1280) & 0.0968 (1120) \\
 \hline
 256 & 0.1392 (200) & 0.1453 (240) & 0.1374 (160) & 0.1513 (160) & 256 & 0.0964 (960) & 0.0991 (800) & 0.1017 (720) & 0.0972 (880) \\
 \hline
 512 & 0.1559 (120) & 0.1694 (160) & 0.1311 (120) & 0.1429 (160) & 512 & 0.0993 (960) & 0.1007 (1040) & 0.1056 (1120) & 0.0953 (1040) \\
 \hline
\end{tabular}
\caption{RL training results for 2D toric code and 2D Ising model with various network size $d_{\rm{NN}}$.
We construct the RL agent with a fully connected neural network with two hidden layers of size $d_{\rm{NN}}$ (for both the policy and value network of a PPO algorithm) and the RL environment with the parameters shown in the table.
In each cell, we record a final reward and the required number of epochs before training termination --- ``Reward (Number of epochs)'' --- for each $d_{\rm{NN}}$ and training index.}

\label{tab:varyNN}
\end{table*}
\renewcommand{\arraystretch}{1}

\begin{table*}[t]
\renewcommand{\arraystretch}{1.5}
\centering
\begin{tabular}{ ||c ||c | c | c | c || c||} 
 \hline
 & RL LEC - Run 1 & RL LEC - Run 2 & RL LEC - Run 3 & RL LEC - Run 4 & Conventional LEC \\
 \hline\hline
 2D Toric Code & 90.62 $\pm$ 0.57 & 90.29 $\pm$ 0.58 & 91.46 $\pm$ 0.55 & 91.54 $\pm$ 0.55 & 86.39 $\pm$ 0.67 \\ 
 \hline
 2D Ising Model & 90.28 $\pm$ 0.58 & 90.64 $\pm$ 0.57 & 89.85 $\pm$ 0.59 & 89.80 $\pm$ 0.59 & 86.48 $\pm$ 0.67\\
 \hline
 4D Toric Code & 95.28 $\pm$ 0.42 & 92.82 $\pm$ 0.51 & 92.47 $\pm$ 0.52 & 93.90 $\pm$ 0.47 & 87.06 $\pm$ 0.66 \\
 \hline
\end{tabular}
\caption{Reward of LEC circuit optimized by 4 RL training executions and conventional LEC circuit. 
The reward is computed out of 10000 independent samples, and the unit of reward is \%. 
Also, we choose $p_{\rm{gate}} = 1\times 10^{-4}$ and $p_{\rm{gate}} = 1\times 10^{-5}$ for 2D and 4D toric code, respectively, and $p_{\rm{gate}} = 1 \times 10^{-3}$ for 2D Ising model (with other parameters as in Table \ref{tab: parameters}).
For all training executions in the same code, we use the same code and error model parameters.}
\label{tab: RL-variations}
\end{table*}
\renewcommand{\arraystretch}{1}

This stability of RL training can be explained by a \emph{simple} state transition, which, in principle, results in low variance in training.
In general, RL is applied to problems where the next state $s'$ is obtained from the conditional probability distribution $P$, given the current state $s$ and action $a$, i.e., $s' \sim P(s' | s, a)$~\cite{10.5555/1622737.1622748, foster2023foundationsreinforcementlearninginteractive}.
For example, in the Gym Acrobot game~\cite{1606.01540}, $s'$ would be updated from $s$ and $a$ based on stochastic laws of physics, which are hidden from the RL agent and have to be ``learned.''
However, in our problem, $s'$ is determined simply by appending $a$ to $s$.
Thus, our RL training is expected to find the optimized solution more reliably compared to the general application of RL methods, which often report substantial training instabilities~\cite{BUSONIU20188, wang2021instabilitiesofflinerlpretrained}.

\section{Error Bar Computation with Data} \label{appendix: error bar}

Each error bar represents a 95\% confidence interval for each parameter.
\begin{itemize}
    \item Final reward (in Fig. \ref{fig:2B_OptimizedDepthLayout}) and final logical error rate for 2D toric code $P_{\rm{L}}$ (in Fig. \ref{fig:4A_ReducingGlobal}):
    
    The final reward is either a decoding success rate (for 2D and 4D toric code) or a ``survival'' rate of an initial spin direction (for the 2D Ising model).
    %
    % $P_{\rm{L}}$ is defined as a decoding failure rate for 2D toric code.
    %
    From large $N$ samples noted in the captions, we compute each of their estimates as a mean of a binomial distribution.
    %
    %Thus, we consider their estimate $\hat{p}$ computed from large $N$ samples in each code as a mean of a binomial distribution with a 95\% confidence interval.
    % \begin{equation}
    %     \hat{p} \pm 1.96 \times \sqrt{\frac{\hat{p} (1- \hat{p})}{N}}.
    % \end{equation}

    \item Average memory lifetime $T$ (in Fig. \ref{fig:3C-1_FittingDeff}):
    
    We compute its estimate as a mean of 1000 samples of independently simulated memory lifetime.
    %
    % \begin{equation}
    %     \text{Mean}_i(T_i) \pm 1.96 \times \frac{\text{Stdev}_i(T_i)}{\sqrt{1000}}.
    % \end{equation}

    \item Effective code distance $D_{\rm{eff}}$ (in Fig. \ref{fig:3C-2_ComparingDeff}) and another fitting parameter $k_1$ (in Fig. \ref{fig:k1-fitting}):

    We obtain fitting parameters and their standard deviations with the symfit library~\cite{Roelfs2017}.
    Note that we set the parameter \boxed{\text{absolute\_sigma}} to be False.
    %
    % Using the symfit librarys with \boxed{\text{absolute\_sigma = False}}~\cite{Roelfs2017}, we fit $D_{\rm{eff}}$ and other parameters and obtain a value $\hat{D}_{\rm{eff}}$ and a standard deviation $\sigma_{D_{\rm{eff}}}$ for each parameter.
    %
    % Then, we compute the 95\% confidence interval for $D_{\rm{eff}}$ (similarly for $k_1$)} as
    % \begin{equation}
    %    \hat{D}_{\rm{eff}} \pm 1.96 \times \sigma_{D_{\rm{eff}}}.
    % \end{equation}
\end{itemize}

\section{Details on memory lifetime extension}

\subsection{Effect of LEC on error statistics} \label{appendix: classification-detail}

% \subsection{Effect of LEC circuits on error statistics}\label{sec: IVA}

% \ptitle{Reward better in RL-optimized LEC}

\begin{figure}[t]
    \centering
    \includegraphics[width = 0.48 \textwidth]{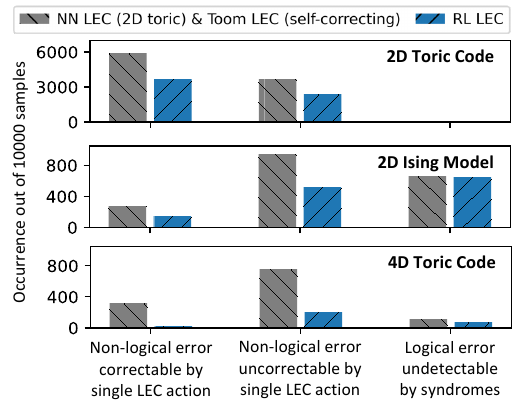}
    \caption{Classification of remaining errors after LEC cycle(s) in each code.
    Let a single LEC action be any LEC action among extended error removal operations as in Fig. \ref{fig:1C_ExtendedOperations}.
    Then, we perform a benchmark simulation for applying LEC cycle(s) with 10000 independent samples.
    Note that for each LEC circuit, $p_{\rm{gate}}$ is $1\times 10^{-4}$ for 2D toric code, $1\times 10^{-3}$ for 2D Ising model, and $1\times 10^{-5}$ for 4D toric code.
    Each error pattern after LEC cycle(s) is either (1) a non-logical \emph{low-weight} error pattern correctable by one of single LEC actions, (2) a non-logical \emph{high-weight} error pattern uncorrectable by any single LEC action, or (3) a logical error \emph{undetectable} by extracting syndromes.
    We compare the occurrence of each category between RL-optimized LEC circuits (blue) vs. conventional LEC circuits in each code (grey).
    %
    % More detailed classification is presented in Appendix \ref{appendix: classification-detail}.
    } 
    \label{fig:3A_ErrorDistribution}
\end{figure}

In Section \ref{sec: IIIB}, we confirm that the RL LEC performs better than the conventional LEC in each code.
%
% Recall that the conventional LEC for 2D toric code is called a nearest-neighbor LEC circuit (or \emph{NN LEC})~\cite{Smith_2023} and that for 2D Ising model and 4D toric code is called a Toom's rule-based LEC circuit (or \emph{Toom LEC})~\cite{10.5555/3179532.3179533}.
%
More specifically, RL LEC shows a lower decoding failure rate compared to conventional LEC after a certain number of repeated LEC cycles.
To analyze the reason for this behavior, we classify remaining error patterns after repeated LEC cycle(s) in each code as shown in Fig. \ref{fig:3A_ErrorDistribution}.
Although both RL and Toom LEC show similar occurrences of logical errors, we observe that RL LEC is effective in suppressing non-logical errors that are either correctable or uncorrectable by a single LEC action that consists of the circuits.
%
% We discuss more details about this classification of remaining errors after repeated LEC cycles in Appendix \ref{appendix: classification-detail}.

% \bigskip\ptitle{Intuitive connection to lifetime}

In each error-correcting code of Fig. \ref{fig:3A_ErrorDistribution}, it is indeed expected that the RL LEC reduces the occurrence of the error patterns correctable by their own individual operations.
However, over repeated cycles of application, the LEC circuits also reduce the accumulation of higher-weight error patterns that are uncorrectable by their own individual operations.
Since these high-weight errors lead to the logical error rate after the final recovery step, \emph{preventing} such errors enable the RL LEC to suppress the logical error rate more effectively compared to the conventional one.

\subsection{Lifetime vs. size for 2D toric code} \label{appendix: T-L-2Dtoric}

In Fig. \ref{fig:3C-1_FittingDeff}, we observe that $T$ slightly decreases as $L$ increases for the RL-optimized LEC circuit.
We conclude that this is because length-$N$ error chains contribute differently to $T$.
As $L$ gets larger, final recovery by perfect syndrome extraction with perfect ancilla readout and MWPM classical algorithm can correct longer error chains.
Thus, the effect of LEC circuits in extending the memory lifetime gets \emph{weaker} due to the varying power of final global decoding as $L$ varies.

\begin{figure}[t]
    \centering
    \includegraphics[width=\linewidth]{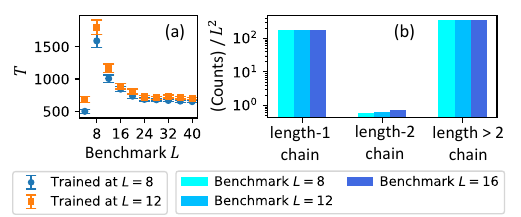}
    \caption{Comparison between data with varying $L$ benchmarked with perfect gates.
    (a) Average memory lifetime $T$ vs. benchmark lattice size $L$ for an LEC circuit trained with $L = 8$ (blue circle) and an LEC circuit trained with $L = 12$ (orange square).
    For both training, $H_{\max} = 40$, $p_{\rm{amb}} = 0.02$, $N = 100$, $R = 7$, and $p_{\rm{gate, train}} = 1\times 10^{-4}$.
    (b) Number of error chains out of 1 million samples divided by $L^2$ for each category of error chain after two LEC rounds.
    For each benchmark $L$ case, we use the common RL-optimized LEC circuit trained at $L = 8$, $H_{\max} = 40$, $p_{\rm{amb}} = 0.02$, $N = 100$, $R = 7$, and $p_{\rm{gate, train}} = 1\times 10^{-4}$.
    Note that logical error occurs in only three samples for the $L=8$ case and none for other $L$s.
    }
    \label{fig:2Dtoric-lifetime-size}
\end{figure}

Fig. \ref{fig:2Dtoric-lifetime-size} supports this explanation.
% \begin{itemize}
In Fig. \ref{fig:2Dtoric-lifetime-size}(a), we find that $T$ decreases as benchmark $L$ increases for both RL-optimized LEC circuits with perfect gates trained at $L = 8$ and $L = 12$, respectively.
Especially, we observe the ``drop'' of lifetime between benchmark $L = 8$ and $L = 12$.
This implies that the relation between a lifetime and lattice size is independent of the training lattice size for optimizing the LEC circuits.
Also, in Fig. \ref{fig:2Dtoric-lifetime-size}(b), we observe that each length-$N$ error chains occur with a similar \emph{density} regardless of benchmark $L$ for a fixed LEC circuit with perfect gates.
This implies that before the final recovery, the effect of the LEC circuit in reducing the number of error chains is similar regardless of benchmark $L$.
Thus, we explain this relation between a lifetime and lattice size based on the role of final recovery.

% \end{itemize}

\subsection{Other parameters from log-log fitting between lifetime and ambient error rate} \label{appendix:  other-lifetime-parameters}

\begin{figure}[t]
    \centering
    \includegraphics[width=\linewidth]{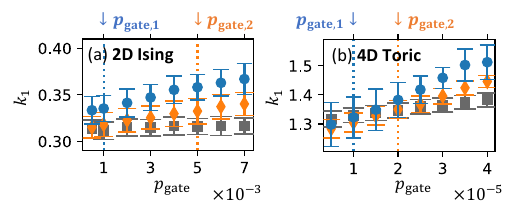}
    \caption{$k_1$ vs. Benchmark $p_{\rm{gate}}$ for RL-optimized LEC circuits trained at $p_{\rm{gate},1}$ (blue circle) and $p_{\rm{gate},2}$ (orange diamond) as well as Toom LEC circuit (grey square) in (a) 2D Ising model and (b) 4D toric code.
    Our computation for the error bar of $k_1$ is discussed in Appendix \ref{appendix: error bar}.}
    \label{fig:k1-fitting}
\end{figure}

Although Section \ref{sec: IVC} focuses on the fitting parameter $D_{\rm{eff}}$ from fitting $\log_{10} T$ and $\log_{10} (p_{\rm{amb}})$, there exist other fitting parameters as well: $k$ for 2D toric code and $k_1$ and $k_2$ for self-correcting memories.
%
% By comparing Equation \eqref{eq: model-2Dtoric} and \eqref{eq: model-self} with \eqref{eq:lifetime_model}, we observe that
% \begin{align}
%     k &= \log_{10} C - D_{\rm{eff}} \times \log_{10} \alpha \\ 
%     k_1 &= \log_{10} \alpha \label{eq: alpha-k_1} \\
%     k_2 &= \log_{10} C
% \end{align}
For 2D Ising model and 4D toric code, we plot the fitted $k_1$ vs. benchmark $p_{\rm{gate}}$ as shown in Fig. \ref{fig:k1-fitting} to check whether its behavior is consistent with the intuition on $\alpha$.
Note that we also confirm from the fitting that the standard deviation for $k$ and $k_2$ do not diverge as well.

Note that $k_1$ shows different values for different $p_{\rm{gate}}$ and LEC circuits.
We expect that $\alpha$ varies for different $p_{\rm{gate}}$ and LEC circuits.
From Equation \eqref{eq: model-self}, increasing $k_1$ implies increasing $\alpha$, because a logarithm function is monotonically increasing.
Physically, $\alpha$ denotes how many new errors \emph{get introduced} every LEC round.
Then, we can explain two general trends observed in Fig. \ref{fig:k1-fitting} based on this intuition on $\alpha$.
% \begin{itemize}

For each LEC circuit, $k_1$ increases as benchmark $p_{\rm{gate}}$ increases.
This trend is consistent with the intuition on $\alpha$ --- as $p_{\rm{gate}}$ increases, more new errors are introduced to the system.
Also, for each benchmark $p_{\rm{gate}}$, $k_1$ is the biggest for the RL-optimized LEC circuit trained with smaller $p_{\rm{gate}}$ and the smallest for the Toom LEC circuit.
Note that in Section \ref{sec: IIIB}, we show that RL-optimized LEC circuit trained with larger $p_{\rm{gate}}$ includes less higher-weight error removal operations. 
%such as $d=1$ and $d=2$ actions for 2D Ising model or $d=1$ and $d=(1,1)$ actions for 4D toric code.
%
This is because the higher-weight error removal operations may introduce more errors into the system compared to the Toom operation. 
%
% Especially, for 4D toric code, we ``grouped'' parallelized $d=1$ and $d=(1,1)$ operations each into single actions (see Appendix \ref{appendix: QECC-implementation}).
%
Thus, for the same benchmark $p_{\rm{gate}}$, $k_1$ is largest for the most ``complicated'' LEC circuits that are trained with smaller $p_{\rm{gate}}$ and smallest for the ``simplest'' Toom circuits.

% \end{itemize}

\begin{figure}[b]
    \centering
    \includegraphics[width=\linewidth]{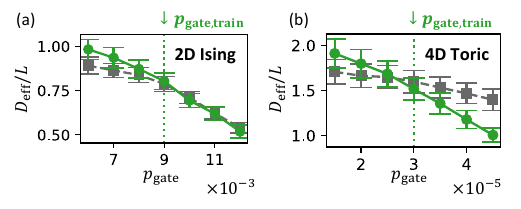}
    \caption{$D_{\rm{eff}}$ vs. Benchmark $p_{\rm{gate}}$ for RL-optimized LEC circuit trained at large $p_{\rm{gate, train}}$ (green circle) as well as Toom LEC circuit (grey square) in (a) 2D Ising model and (b) 4D toric code.
    Here, we choose $p_{\rm{gate, train}}$ to be $9 \times 10^{-3}$ for 2D Ising model and $3 \times 10^{-5}$ for 4D toric code.}
    \label{fig:large-pgate}
\end{figure}

\subsection{LEC circuits with large gate error rate} \label{appendix: large-gate-error}

For 2D toric code, RL training with large $p_{\rm{gate}}$ results in the simplest NN decoder.
Note that we fix our first action in the LEC circuit to be a syndrome extraction (see Appendix \ref{appendix: RL-agent}).
Thus, we do not observe that the RL framework produces a ``do nothing'' LEC circuit.
However, we find that for $p_{\rm{gate}} > 0.01$, the simplest NN circuit performs worse than just ``do nothing.''

On the other hand, as shown in Fig. \ref{fig:large-pgate}, we observe that the RL-optimized LEC circuit trained at $p_{\rm{gate, train}}$ performs as much as --- or even worse than --- the Toom LEC circuit when benchmark $p_{\rm{gate}}$ is $p_{\rm{gate, train}}$ if $p_{\rm{gate, train}}$ is large enough.
In other words, we observe that $p_{\rm{gate,c}} \approx p_{\rm{gate, train}}$ or even $p_{\rm{gate},c} < p_{\rm{gate, train}}$ for large $p_{\rm{gate, train}}$ case, unlike as shown in Fig. \ref{fig:3C-2_ComparingDeff}.
This feature is the limitation of our RL framework, and this is the reason why we do not investigate improving the threshold gate error rate of Toom's rule decoder using our RL framework.

\section{Details on LEC Application} \label{appendix: min-N-global}

\begin{figure}[b]
    \centering
    \includegraphics[width=\linewidth]{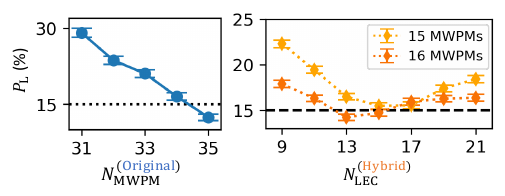}
    \caption{Method of finding minimum number of global decoding for original vs. hybrid decoding protocol to achieve $P_{\rm{L}}\ < 15$\% for fixed $T = 2000$ with 5000 samples of 2D toric code lattice with $L = 16$.}
    \label{fig:minimum-method}
\end{figure}

From our choice of the error model, fidelity of CNOT gate $F_{2\rm{Q}} = 1 - 4\times p_{\rm{gate}}$, and fidelity of CCX/CCZ gate $F_{3\rm{Q}} = 1 - 6\times p_{\rm{gate}}$.
Thus, multi-qubit gates with $p_{\rm{gate}} = 1 \times 10^{-4}$ have $F_{\rm{2Q}} = 99.96\%$ and $F_{\rm{3Q}} = 99.94\%$.
In this regime, we investigate what is the minimum number of global MWPM decoding for original vs. hybrid decoding protocol to achieve the final logical error rate after total time $T$ to be below $15\%$.

% \bigskip\ptitle{Method of optimization}

% We focus on the system including , where the fidelity of the CNOT gate is $99.96\%$ and that of the CCX/CCZ gate is $99.94\%$.
%
% Note that from our choice of the error model, 
%
% In this regime, we investigate what is the minimum number of global decoding for each decoding protocol to achieve the final logical error rate after total time $T$ to be below $15\%$.
%
As shown in Fig. \ref{fig:minimum-method}, we perform 1D and 2D optimization for the original and hybrid decoding protocol, respectively.
For the original decoding, we vary $N_{\rm{MWPM}}$ and check the final $P_{\rm{L}}$ after time $T$.
On the other hand, for the hybrid decoding, we first fix a total number of global decoding and then find $N_{\rm{LEC}}$ that minimizes $P_{\rm{L}}$ after time $T$.
Then, we choose the minimum number of global MWPM decoding for each of the protocols that satisfy $P_{\rm{L}} < 15\%$.

On top of the minimum $N_{\rm{MWPM}}$ for $p_{\rm{gate}} = 1 \times 10^{-4}$ discussed in Section \ref{sec: VB}, we also compare minimum $N_{\rm{MWPM}}$ for larger $p_{\rm{gate}}$ as well.
When $p_{\rm{gate}} = 3\times 10^{-3}$, i.e. $F_{2\rm{Q}} = 99.88\%$ and $F_{3\rm{Q}} = 99.82\%$, we obtain that the hybrid protocol can reduce the number of global decoding by 1 or 2 compared to the original protocol regardless of $T$.
Also, note that we observe the same results for these comparisons, even though the original decoding protocol is improved by replacing the 3D MWPM following repeated faulty syndrome extractions with the 2D MWPM following a perfect syndrome extraction.

\clearpage

\bibliography{bibliography.bib}

\clearpage

\end{document}